# Trade-offs between carbon stocks and biodiversity in European temperate forests

Sabatini* F.M., de Andrade R.B., Paillet Y., Ódor P., Bouget C., Campagnaro T., Gosselin F., Janssen P., Mattioli W., Nascimbene J., Sitzia T., Kuemmerle T., Burrascano S.

## Author affiliations and ORCID:

Sabatini, Francesco Maria. 1) Geography Department, Humboldt-Universität zu Berlin. Unter den Linden 6, 10099 Berlin, Germany. 2) German Centre for Integrative Biodiversity Research (iDiv) - Halle-Jena-Leipzig; 3) - Martin-Luther-Universität Halle-Wittenberg, Institut für Biologie, Am Kirchtor 1, 06108, Halle (Saale). Germany. ORCID: 0000-0002-7202-7697. * Corresponding author:

francesco.sabatini@botanik.uni-halle.de

de Andrade, Rafael Barreto. Sapienza, University of Rome. P.le Aldo Moro 5, 00165 Rome, Italy. ORCID: http://orcid.org/0000-0002-7976-7647

Paillet, Yoan. Irstea, UR EFNO, Domaine des Barres, 45290 Nogent sur Vernisson, France. ORCID: 0000-0001-7232-7844

Ódor, Peter. MTA Centre for Ecological Research, Institute of Ecology and Botany, H-2163 Vácrátót, Alkotmány u. 2-4., Hungary. ORCID: 0000-0003-1729-8897

Bouget, Christophe. Irstea, UR EFNO, Domaine des Barres, 45290 Nogent sur Vernisson, France. ORCID: 0000-0002-5206-7560

Campagnaro, Thomas, Department of Land, Environment, Agriculture and Forestry, Università degli Studi di Padova, I-35020 Legnaro (PD), Italy. thomas.campagnaro@unipd.it ORCID: 0000-0002-6307-6892

Gosselin, Frédéric. Irstea, UR EFNO, Domaine des Barres, 45290 Nogent sur Vernisson, France. ORCID : 0000-0003-3737-106X

Janssen, Philippe. Université Grenoble-Alpes, Irstea, UR LESSEM, 2 rue de la Papeterie BP 76, 38402 Saint-Martin-d'Hères, France. ORCID: 0000-0003-3310-0078

Mattioli, Walter. Council for Agricultural Research and Economics, Research Centre for Forestry and Wood (CREA-FL), v.le Santa Margherita 80, I-52100 Arezzo (AR), Italy, walter.mattioli@crea.gov.it. ORCID: 0000-0003-2107-1038

Nascimbene, Juri. University of Bologna, Department of Biological, Geological and Environmental Sciences, via Irnerio 42, 40126 Bologna. ORCID: 0000-0002-9174-654X.

Sitzia, Tommaso. Department of Land, Environment, Agriculture and Forestry, Università degli Studi di Padova, I-35020 Legnaro (PD), Italy. tommaso.sitzia@unipd.it. ORCID: 0000-0001-6221-4256

Kuemmerle, Tobias. 1) Geography Department, Humboldt-Universität zu Berlin. Unter den Linden 6, 10099 Berlin, Germany 2) Integrative Research Institute for Human Environment Transformation (IRI THESys), Humboldt-Universität zu Berlin. Unter den Linden 6, 10099 Berlin, Germany.  ORCID: 0000-0002-9775-142X

Burrascano, Sabina. Sapienza, University of Rome. P.le Aldo Moro 5, 00165 Rome, Italy. ORCID: http://orcid.org/0000-0002-6537-3313

## Abstract

Policies to mitigate climate change and biodiversity loss often assume that protecting carbon-rich forests provides co-benefits in terms of biodiversity, due to the spatial congruence of carbon stocks and biodiversity at biogeographic scales. However, it remains unclear whether this holds at the scales relevant for management, with particularly large knowledge gaps for temperate forests and for taxa other than trees. We built a comprehensive dataset of Central European temperate forest structure and multi-taxonomic diversity (beetles, birds, bryophytes, fungi, lichens, and plants) across 352 plots. We used Boosted Regression Trees to assess the relationship between above-ground live carbon stocks and (a) taxon-specific richness, (b) a unified multidiversity index. We used Threshold Indicator Taxa ANalysis to explore individual species' responses to changing above-ground carbon stocks and to detect change-points in species composition along the carbon-stock gradient. Our results reveal an overall weak and highly variable relationship between richness and carbon stock at the stand scale, both for individual taxonomic groups and for multidiversity. Similarly, the proportion of win-win and trade-off species (i.e. species that respectively increase or decrease with increasing carbon) varied substantially across taxa. Win-win species gradually replaced trade-off species with increasing carbon, without clear thresholds along the above-ground carbon gradient, suggesting that aggregate indices (e.g. richness) might fail to detect these changes. Collectively, our analyses highlight that leveraging co-benefits between carbon and biodiversity in temperate forest may require stand-scale management that prioritizes either biodiversity conservation or carbon-storage in order to maximize co-benefits at broader scales. Importantly, this contrasts from tropical forests, where climate and biodiversity objectives may be effectively integrated at the stand-scale, thus highlighting the need for context-specificity when managing for multiple objectives. Accounting for critical change-points of target taxa can help to deliver this specificity, by defining a safe operating space to manipulate carbon while avoiding biodiversity losses.

## Introduction

Forests play a critical role in mitigating climate change, in addition to providing many ecosystem services fundamental to human society (FAO, 2015; MEA, 2005). The estimated amount of carbon stored in forests globally is almost 900 Pg (=$10^{15}$ g), with a net global carbon sink of 1.1 Pg C per year (Pan et al., 2011). Forests also provide habitat for over half of all known terrestrial plant and animal species (MEA, 2005), albeit covering only 27% of the Earth's land area (FAO, 2015). Conserving forests and managing them sustainably is therefore fundamental for facing two of the most pressing societal challenges of our times: biodiversity loss and climate change (MEA, 2005).

Global and regional environmental policies, such as the 2015 Paris Agreement, the REDD+ (Reducing Emissions from Deforestation and forest Degradation) initiative (Gardner et al., 2012) or the European Forest Strategy (European Commission, 2013), all acknowledge the critical importance of forests for jointly addressing biodiversity conservation and climate change mitigation (Bustamante et al., 2016; Deere et al., 2018; Ferreira et al., 2018). The extent to which these two targets can be reached in parallel, however, is not properly understood (Di Marco, Watson, Currie, Possingham, & Venter, 2018; Mori, Lertzman, & Gustafsson, 2017; Pichancourt, Firn, Chadès, & Martin, 2014). If high biodiversity and carbon stocks coincide spatially, then protecting carbon-dense forests or managing forests for high carbon stocks would co-benefit both environmental policy goals (Di Marco et al., 2018; Reside, VanDerWal, & Moran, 2017; Strassburg et al., 2010). Otherwise, this may lead to negative biodiversity outcomes (Boysen, Lucht, & Gerten, 2017; Bustamante et al., 2016; Ferreira et al., 2018). For instance, protecting a carbon-dense forest may reallocate human pressure to unprotected areas with lower carbon density, but high biodiversity (Di Marco et al., 2018). Also, shifting from natural vegetation to tree plantations to maximise carbon stock leads to biodiversity loss (Pichancourt et al., 2014), especially where natural grasslands or savannahs are afforested (Bremer & Farley, 2010; Burrascano et al., 2016; Pellegrini, Socolar, Elsen, & Giam, 2016). Finally, it remains unclear at which scales co-benefits between conservation and climate-change-mitigation should be sought. Should carbon storage and conservation goals be integrated at the stand scale, where management takes place, or at broader scales, thus prioritizing one of the goals at the stand scale?

Understand the relationship between carbon stocks and biodiversity, and how it varies across spatial scales, is crucial to answer this question (Gardner et al., 2012; Mori et al., 2017; Reside et al., 2017).

At the scale of an individual forest stand, carbon stock is the amount of long-term carbon stored in living biomass, dead organic matter and soil carbon pools, and is the result of the complex relationships between forest productivity, disturbance history, and species composition (FAO, 2015). Compared to other important forest carbon pools, above-ground live carbon stored in wood (hereafter above-ground live carbon) can be quantified relatively easily, and is therefore considered a sustainable forest management indicator (CBD, 2006; FOREST EUROPE, 2015). Even if it constitutes the substrate for many forest species (Bouvet et al., 2016; Hatanaka, Wright, Loyn, & Mac Nally, 2011; Stokland, Siitonen, & Jonsson, 2012), above-ground live carbon is only indirectly related to biodiversity (Hatanaka et al., 2011). Recent evidence supports a positive correlation between above-ground live carbon and biodiversity at broader scales (Di Marco et al., 2018; Strassburg et al., 2010). The shape of the carbon-biodiversity relationship remains however an open question for the scales most relevant to decision-makers, such as landscape and stand scales (Deere et al., 2018; Ferreira et al., 2018; Pichancourt et al., 2014). Large uncertainties also remain on how this relationship varies across biogeographical regions (Di Marco et al., 2018; Potter & Woodall, 2014; Xian et al., 2015). For the tropics, there is evidence for a positive relationship between biodiversity and above-ground live carbon stocks, both across stands (Cavanaugh et al., 2014; Deere et al., 2018; Magnago et al., 2015), and within stands (Sullivan et al., 2017), especially for disturbed sites (Ferreira et al., 2018). In temperate forests, research has traditionally focused on carbon sequestration and productivity (Ratcliffe et al., 2017), which are often not good indicators of carbon stock. Although recent work highlighted the importance of carbon quality, or complexity, for bird biodiversity (Hatanaka et al., 2011), studies relating carbon quantity to biodiversity remain rare and mostly refer to tree diversity only (Potter & Woodall, 2014; Xian et al., 2015).

The carbon-biodiversity relationship may also vary across taxonomic groups (Di Marco et al., 2018; Ferreira et al., 2018). Even in the tropics most research to date has focused on either vertebrates (Beaudrot et al., 2016; Deere et al., 2018; Sollmann et al., 2017), or tree species richness only (Cavanaugh et al., 2014;

Magnago et al., 2015; Sullivan et al., 2017), while research comparing the carbon-biodiversity relationship across groups at fine-scale remains rare (Ferreira et al., 2018). Although understandable, given the inherent costs of collecting field-based data for multiple taxonomic groups (Bustamante et al., 2016), focusing on trees or vertebrates only assumes that these taxa are good surrogates for overall forest biodiversity, which is questionable (Larrieu et al., 2018; Sabatini et al., 2016; Zilliox & Gosselin, 2014). Moreover, even within the same taxonomic group, different species may relate very differently to carbon stocks (Edwards et al., 2014; Lindenmayer, Fischer, & Cunningham, 2005; Villard & Jonsson, 2009). Some species may benefit from increasing carbon stocks (hereinafter called 'win-win species'), while others, hereafter 'trade-off species', may be hindered by the environmental conditions associated to carbon-dense forests (Ferreira et al., 2018). Splitting the community into win-win and trade-off species, and considering explicitly the behaviour of species of conservation concern, could thus help to better predict the effect of changing carbon stock on specific components of biodiversity (Magnago et al., 2015; Sollmann et al., 2017).

Finally, although the carbon-biodiversity relationship is often assumed to be linear (Beaudrot et al., 2016; Deere et al., 2018; Sullivan et al., 2017), thresholds could exist along the carbon stock gradient, meaning that a slight change in forest carbon stocks could cause disproportionate biodiversity loss (Evans et al., 2017; Sasaki, Furukawa, Iwasaki, Seto, & Mori, 2015). Such thresholds have been identified for a range of anthropogenic gradients (Li, Xu, Zheng, Taube, & Bai, 2017; Magnago et al., 2015; Sasaki et al., 2015), including carbon stocks in tropical forests (Ferreira et al., 2018). For temperate forests, however, empirical evidence is lacking (Evans et al., 2017), especially at fine scales (Sasaki et al., 2015). Identifying such thresholds, and understanding how they vary across taxa and forest types, would provide important information on how forest management, including timber harvesting, might impact biodiversity. This would help to identify 'safe operating spaces' for manipulating forest carbon in managed forests without triggering undesired biodiversity loss (Villard & Jonsson, 2009).

Here, we investigated the relationships between the diversity of six ecological groups (i.e., saproxylic beetles, birds, bryophytes, wood-inhabiting fungi, epiphytic lichens, and vascular plants) and carbon stock across 22 temperate forest sites in three European countries. We addressed the following questions:

(1) What is the relationship between above-ground live carbon stocks and (a) species richness of different taxa, and (b) a single, unified multidiversity index?

(2) How do responses to increasing above-ground live carbon of individual species, and the proportion of win-win and trade-off species, vary across taxonomic groups and forest types?

(3) Are there community level thresholds in species richness or composition along carbon stock gradients?

## Methods

### Study sites

Our study area included a network of 352 plots in 22 temperate forest sites (ranging from 200 to 400 km$^2$) covering a wide latitudinal and longitudinal range across Europe (Figure 1, Table S1). The sites covered deciduous forest types that are common in temperate Europe, including acidophilous oak and oak-birch forest (20 plots), mesophytic deciduous forest (84 plots), European beech (*Fagus sylvatica*) and montane beech forest (232 plots in total), as well as thermophilous deciduos forest (16 plots). Forest type nomenclature follows EEA (2006). Our dataset also covered a wide range of structural types (one-, two- and multi-layered stands), ages, management histories, and management regimes, including coppice, shelterwood, group selection and unmanaged stands. These stands comprised most of the forest succession gradient, including late-successional phases.

Elevation ranged from 150 to 1700 m a.s.l. and substrates included sedimentary rocks (limestones, dolomites, marls, and flysch) in the French and Italian sites, and alluvial gravel mixed with sand and loess in the Hungarian site. All sites were within the temperate region: annual mean temperature varied from 5°C in the French Alps to 14°C in southern Italy. Annual precipitation varied from about 600 mm to about 1900 mm.

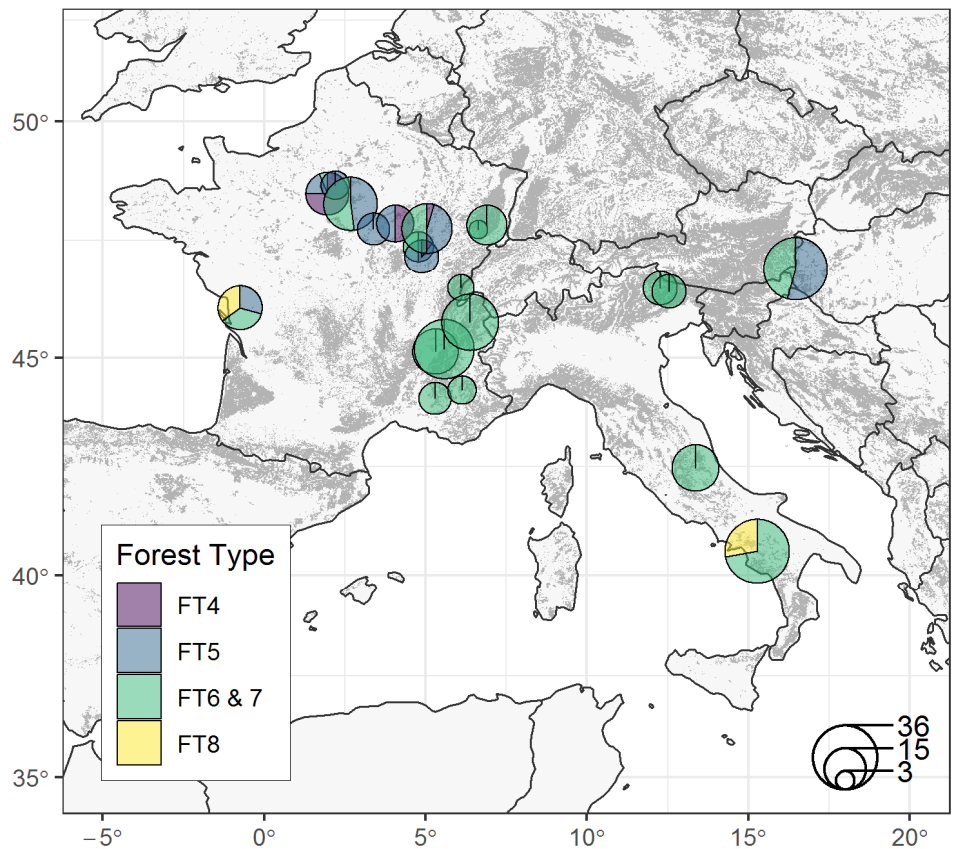

*Figure 1 – Distribution of forest sites in Europe. Pie charts return the relative proportion of plots in different forest types (FTs) for each stand. FTs follow EEA (2006). The size of the pie represents the number of plots in each site. Grey shadings represent the distribution of forest in Europe. FT4 - acidophilous oak and oak-birch forest; FT5 - mesophytic deciduous forest; FT6 - European beech (Fagus sylvatica) and (FT7) montane beech forest; FT8 - thermophilous deciduos forest.*

### Sampling and measuring biodiversity

Our multi-taxonomic dataset stemmed from projects (Burrascano et al., 2018) using similar, but not identical sampling protocols (Table S2 in the Supplementary Material for details). All vascular plant species were recorded in plots ranging from 314 to 1,256 m$^2$. Bryophytes were sampled on different substrates (standing trees, deadwood, rocks, and ground), while only epiphytic lichens and wood-inhabiting fungi were considered. Birds were sampled through point-counts or area search techniques. Saproxylic beetles were sampled using window-flight interception traps, emerging traps and Winkler extractors. Only presence-absence data was available for birds, bryophytes, fungi and lichens, while abundance data was available for plants (percentage cover) and beetles (number of trapped individuals). Not all six taxonomic groups were sampled in every plot, returning a total of 1,533 (= taxonomic group x plot) combinations. We fixed nomenclature inconsistencies in the species lists based on up-to-date checklists (Table S2).

To control for varying sampling efforts across sites, we calculated for each plot the scaled richness of each taxonomic group, i.e. the ratio between the richness observed in the plot (alpha diversity) and the species pool size of that taxonomic group in a given site. Species pool size was estimated as the asymptotic species richness based on sample-based rarefaction and extrapolation curves (Colwell et al., 2012) using the Chao2 estimator in the R package *iNext* (Hsieh, Ma, & Chao, 2016). We then calculated the average scaled richness across taxonomic groups to obtain a single measure of the diversity of all taxa we sampled, hereafter referred to as multidiversity (Allan et al., 2014). Multidiversity ranges between 0 and 1, with multidiversity of 1 meaning that a plot hosts all species contained in the species pool of a site. Multidiversity has the advantage of being comparable across sites, whatever the sampling effort and the species pool (both at the taxa and the site levels).

### Forest above-ground live carbon

We sampled living trees in plots ranging from 491 and 2,827 $m^2$ in area using a diameter at breast height (DBH) threshold of 10 cm. Height was measured for all the trees or in a sample of them and calculated successively for the others by means of height-diameter models. Growing stock was calculated using local allometric equations, with DBH and height as explanatory variables (Table S2) and then converted to above-ground live carbon (AGC, MgC/ha) as AGC = GS*BEF*WBD/2, where BEFs are biomass expansion factors and WBDs are the wood basic densities (Federici, Vitullo, Tulipano, De Lauretis, & Seufert, 2008).

### Control variables

Coarse woody debris was sampled in plots ranging from 491 to 1,600 $m^2$ using a diameter threshold of 10 cm. Volume of deadwood pieces were either calculated using the same allometric equations used for living trees (for standing or downed dead trees), or approximating deadwood volume to truncated cones or cylinders, depending on the data source (Table S2). We then calculated coarse woody debris ratio as the ratio between deadwood volume and total live and dead wood volumes. As additional control variables, we derived two topographic covariates (slope, aspect) from a 30-m resolution Digital Terrain Model (NASA, 2006), which we then used to calculate heat load, i.e. the heat gain from incoming solar radiation. Moreover, we derived two climatic variables (mean annual temperature and precipitation) from raster layers with 30 arcsec resolution grids (approx. 13 km) from WorldClim v2.0 (Fick & Hijmans, 2017). We derived information on parent material from the European Digital Archive on Soil Maps to classify plots into three classes: igneous-metamorphic, sedimentary-clastic, sedimentary-limestone (Panagos, Jones, Bosco, & Kumar, 2011). Finally, we considered forest type as a categorical variable with four levels: acidophilous oak, mesophytic deciduous, beech dominated, and thermophilous deciduous.

## Modelling the response of biodiversity to forest above-ground C

We used Boosted Regression Trees (BRTs) to assess the relationship between above-ground live carbon and the scaled species richness of each taxonomic group as well as multidiversity. BRTs are non-parametric models based on decision trees in a boosting framework that does not require prior assumptions. BRTs are therefore relatively robust against overfitting, missing data, and collinearity (Elith, Leathwick, & Hastie, 2008). We used above-ground live carbon as explanatory variable, while controlling for the effect of forest structure, climatic, topographical and soil. After checking for correlation (Pearson's r>0.7), we retained five control variables: coarse woody debris ratio, forest type, annual precipitation, heat load and substrate parent material. We also included *forest site* and *data source* as categorical control variables to account for remaining unobserved environmental and methodological differences across sites.

We parametrized the BRTs setting a tree complexity of 5 and a bag fraction of 0.5 (Elith et al., 2008). We tested different learning rates (0.5 - 0.0025), and determined the optimal number of trees for each learning rate using the *gbm.step* routine provided in the *dismo* package (Hijmans, Phillips, Leathwick, & Elith, 2011). We then selected the parameter combination returning the highest cross-validated model fit. We finally calculated the relative importance of each explanatory variable (i.e., the fraction of times a variable was selected for splitting a tree in each BRT model, weighted by the squared model improvement). We evaluated model performance using 10-fold cross-validation. We explored the relationship between biodiversity and the explanatory variables using partial dependency plots, which are the graphical visualizations of the marginal effect of a given explanatory variable on scaled richness (or multidiversity). These plots also allow to visually check for non-linear responses and possible thresholds. We explored the interactions between explanatory variables using the *gbm.interactions* function in the *dismo* package. All analysis were performed in R 3.4.1.

### Assessment of win-win and trade-offs species

We used Threshold Indicator Taxa Analysis – TITAN (Baker & King, 2010) to identify win-win and trade-off species, i.e. species that respectively increase or decrease their abundance and/or frequency with increasing levels of above-ground live carbon. TITAN uses binary partitioning by indicator value (*IndVal,* Dufrêne & Legendre, 1997) to identify species-specific change-points along an environmental gradient (above-ground live carbon in our case). Change-points are compared to random data permutations to assess their relevance, taking into account indices of purity (i.e. proportion of bootstrapped change-points response that agree with the observed response) and reliability (i.e. proportion of bootstrapped change points with significant *IndVal* for $p<0.05$). We evaluated uncertainty in change-point location based on the bootstrapped empirical distribution (Baker & King, 2010). To account for the nested (plots within sites) and unbalanced (different number of plots per site) nature of our dataset, we modified TITAN's original bootstrapping approach to randomly select (with replacement) a number of plots per site equal to the number of plots in the site having the lowest number of plots (if greater than 3, 3 otherwise). We ran TITAN after pooling all species across taxonomic groups, but separately for forest types. We aggregated acidophilous and mesophytic oak forests (oak-dominated thereafter), and lowland and montane beech forests (beech-dominated forest thereafter) but excluded termophilous oak forests due to the low sample size (n = 16). We also checked the conservation status of our win-win or trade-off species using the IUCN red lists (IUCN, 2017) and the r package *rredlist* (version 4.0, Chamberlain, 2017).

TITAN also allows to explore if species-level change points aggregate to a community-level threshold, i.e. congruent change-points across all individual species, which we did separately for trade-off and win-win species. To explore the variability across taxonomic groups in community-level change-points, we ran TITAN both when considering the whole species assemblage, and for each taxonomic groups. We considered narrow confidence limits across bootstrapped replicates as an evidence for a community threshold.

## Results

### Relationships between above-ground live carbon and species richness

The relationship between scaled richness and above-ground live carbon was overall consistently weak (relative importance between 3.3-12.2%), and varied in direction across taxa (Figure 2). The scaled richness of birds, bryophytes and fungi increased slightly, but non-linearly, across the above-ground live carbon gradient (~5% absolute increase along the whole gradient). Most of the increase for birds and bryophytes occurred between 180-200 MgC/ha and 120-150 MgC/ha, respectively. The richness fraction of fungi increased non-monotonically along the above-ground live carbon gradient, with a first peak at 70 MgC/ha, and a secondary peak at 175 MgC/ha. The richness fraction of plants, instead, showed a ~5% absolute decline (see also Figures S1-S6).

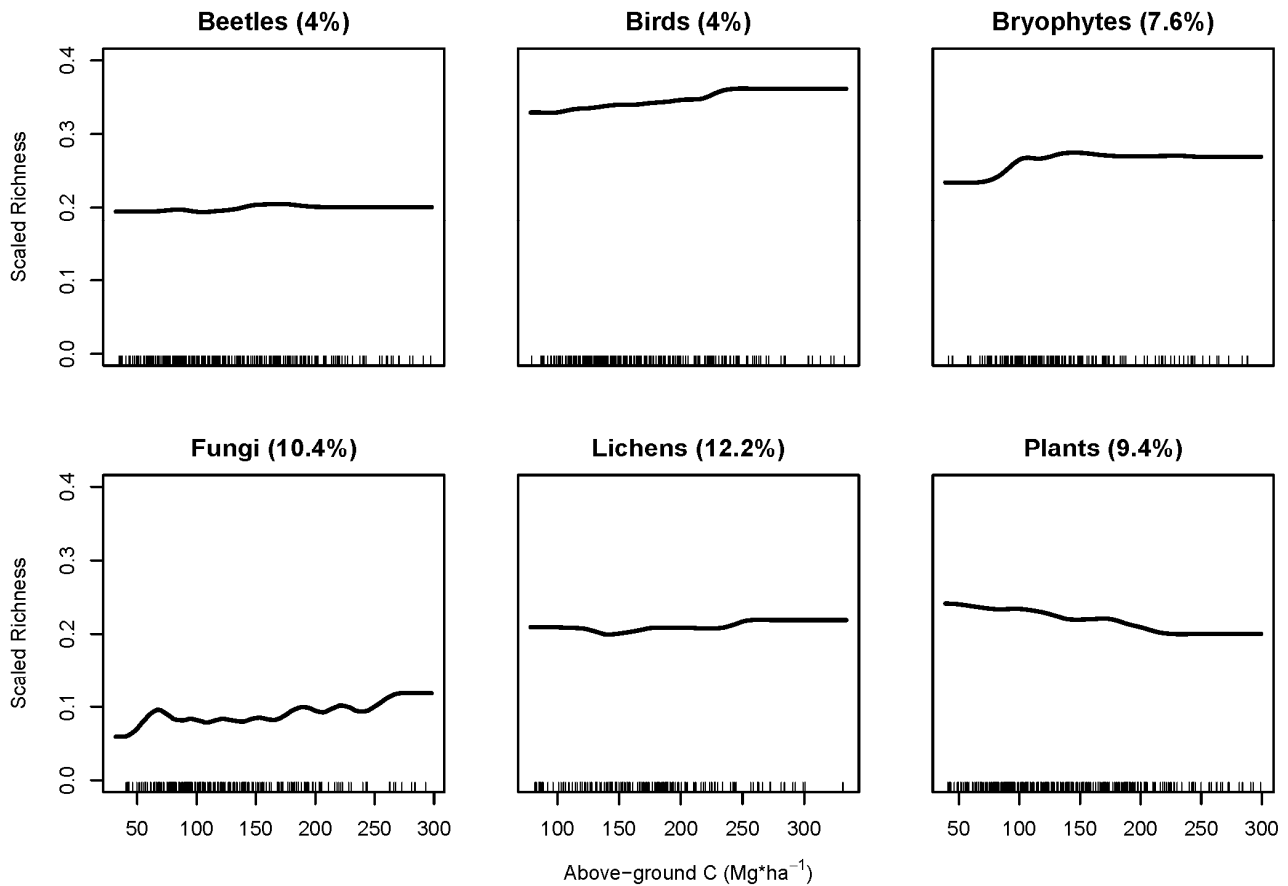

*Figure 2 – Partial dependency plots of the relationship between scaled richness and above-ground live carbon, modelled using Boosted Regression Trees. Scaled richness represents the fraction of species of the species pool size estimated for a given plot. Ticks on the x-axis represent above-ground live carbon data distribution. For each taxonomic group, we report in parenthesis the relative importance of above-ground live carbon in the respective boosted regression tree model.*

Above-ground live carbon had a very little relative importance on multidiversity (3.3%), compared to other control variables especially *site* (relative importance 77.2% - Figure S7). Multidiversity increased by less than 1% over the whole above-ground live carbon gradient (Figure 3). BRT models were effective at modelling multidiversity and scaled richness (cross-validation correlation 0.53-0.84, Table S3), although most of the variation derived from site-to-site differences. Indeed, *site* was always the variable having the highest relative importance (57.4-84%, Figure S1-S7). For all taxa but beetles and lichens, the interaction between *site* and above-ground live carbon ranked among the top three most important interactions.

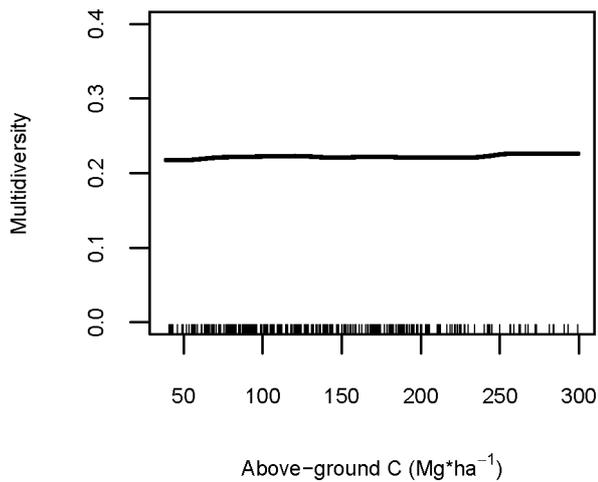

*Figure 3 – Partial dependency plot of the relationship between multidiversity and above-ground live carbon using Boosted Regression Trees. Tick marks on the x-axis represent above-ground live carbon data distribution.*

Response of individual species to changes in above ground carbon

The TITAN analyses identified 27 and 75 species as pure and reliable indicators of above-ground live carbon (i.e. having a consistent response in both direction and magnitude across bootstrap replicates), corresponding to 5.3% and 9.0% of the total number of species in oak- and beech-dominated forests, respectively (Figure 4). Most of the species-specific change-points occurred between 80-120 MgC/ha, in both forest types (Fig. S8-S9). Eleven species (ten plants and one bryophyte) were pure and reliable indicators in both forest types. All pure and reliable trade-off indicator species for oak-dominated forests were plants (Figure S8), either tree species with good dispersal ability (e.g., *Sorbus aria*, *S. domestica*, *Acer campestre*) or herbs and shrubs associated to forest margins (e.g., *Vicia sepium*, *Lonicera xylosteum*, *Rosa arvensis*). Win-win species were mostly bryophytes, typically found in shaded conditions and, secondarily, beetles *(Tomicus piniperda, Cryptolestes duplicatus)*. Trade-off species in beech forests were principally plants (22 species), beetles (16 species) and secondarily birds (three species) and fungi (one species, Figure S9). Win-win species were mostly plants (14 species) and beetles (11 species). Those associated to the right end of the above-ground C gradient were mostly fungi (*Fuscoporia ferruginosa*, *Stereum rugosum* and

*Heterobasidion annosum*) and beetles (e.g., *Pediacus dermestoides*, and *Xylechinus pilosus*). Only 200 out of the 2,384 species we considered were included in the IUCN database, with conservation status of win-win and trade-off species only available for 18 species, none of which were threatened.

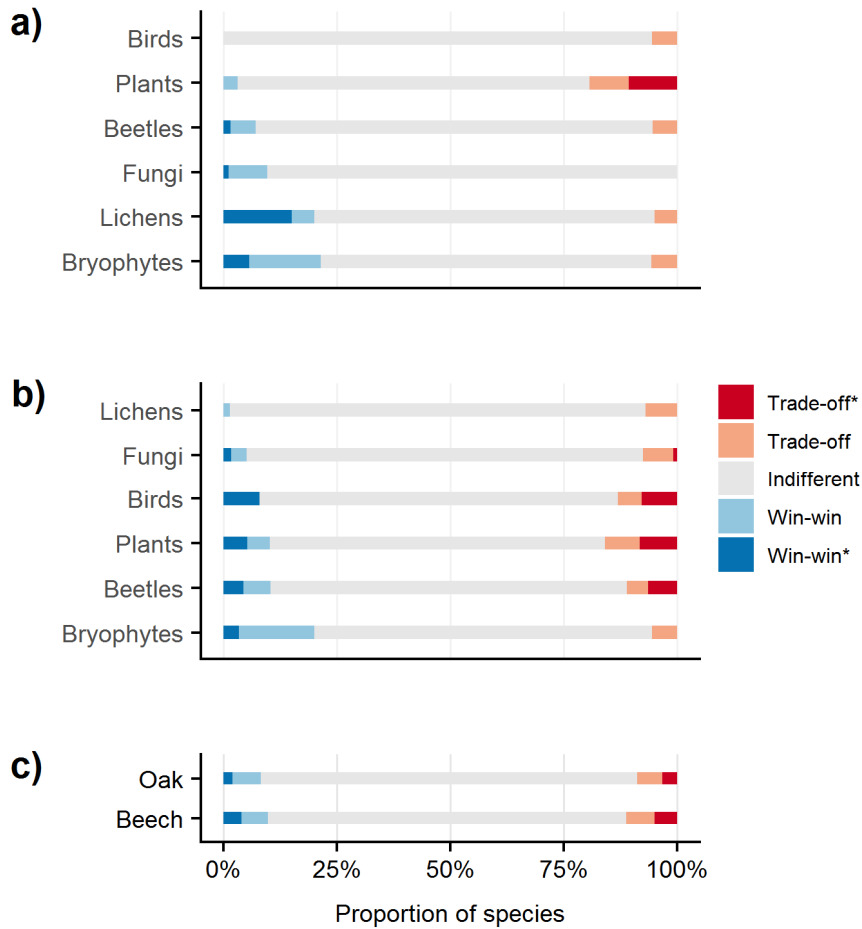

*Figure 4 – Proportion of win-win vs. trade-off species across taxonomic groups and forest types (sorted for decreasing number of win-win species). a) oak-dominated forests. b) beech-dominated forests c) comparison of the two forest types across all taxonomic groups. Win-Win\* (trade-off\*) species are pure and reliable species. Win-win (trade-off) species are pure but not reliable indicators.*

We found a slightly higher number of trade-off than win-win species, both in oak (3.4 vs. 2%) and in beech (4.7% vs. 4.3%, respectively - Figure 4c). When also considering species not having a reliable response (i.e. responding consistently across bootstraps, but being significant indicators at the $p<0.05$ level in less than 95% of bootstraps) this balanced picture did not change (8.7% trade-off vs 7.9% win-win species in oak-dominated forests; 11.3% vs 10.8% in beech dominated forests).

The contribution of individual taxonomic groups varied substantially across forest types. In oak-dominated forests (Figure 4a), fungi, lichens and bryophytes returned a higher proportion of win-win than trade-off species, while for plants, we observed the opposite. In beech forests, most taxonomic groups had a higher proportion of trade-off than win-win species, with the exception of bryophytes and beetles (Figure 4b).

## Community level change-points along above-ground live carbon gradients

Aggregating individual species' responses to infer community-level change-points did not reveal a clear community-level threshold in above-ground live carbon across all the taxa (Figure 5). In oak-dominated forests, the wide confidence intervals around the community-level change-points suggest that rather than abruptly, trade-off species were gradually replaced by win-win species with increasing above-ground live carbon. In beech forests, instead, we observed relatively sharp community-level change-points for trade-off species across all taxa (except lichens), which ranged between 81.3 MgC/ha (fungi) and 122.4 MgC/ha (lichens). In both forest types, community-level change-points of win-win species were more variable than those of trade-off species and, at least in beech forests, returned wider confidence intervals (e.g., for plants, fungi and beetles). Community-level change-points for trade-off species of different taxa were very similar across the two forest types, while for win-win species these were on average higher in beech compared to oak forests.

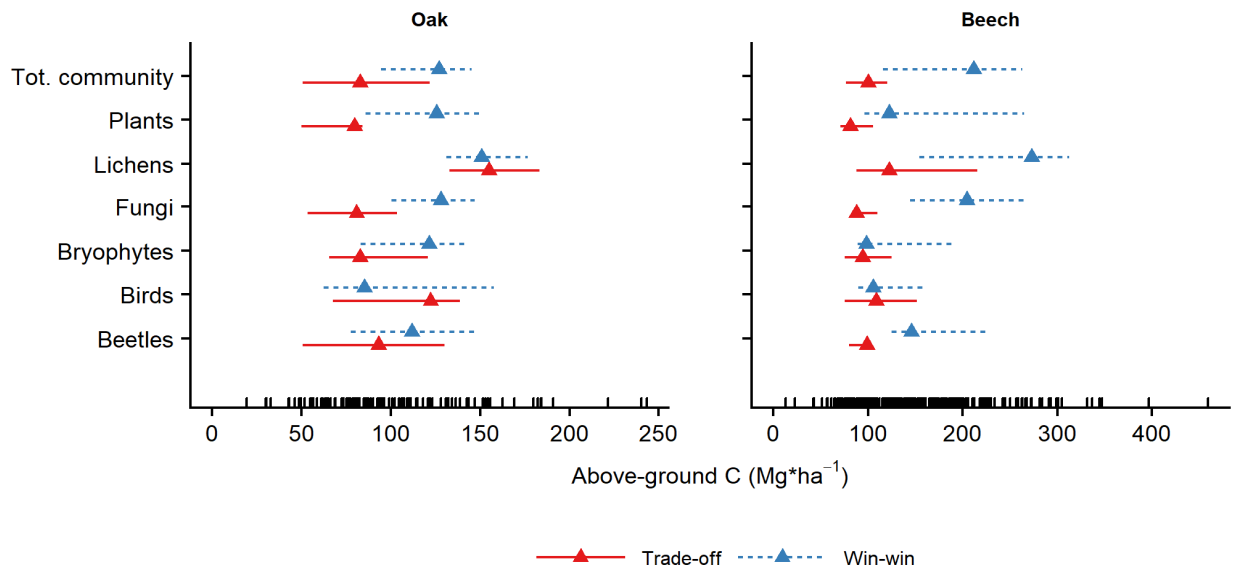

*Figure 5 – Community-level change-points and 90% quantiles along the above-ground live carbon gradient for different taxonomic groups, in two forest types. Tick marks on the x-axis represent above-ground live carbon data distribution.*

## Discussion

Can managing forest for carbon storage jointly achieve biodiversity conservation and climate-change mitigation goals? Answering this question critically depends on better understanding the relationship between forest carbon stocks and biodiversity for the scales at which management takes place (Ferreira et al., 2018; Gardner et al., 2012; Mori et al., 2017). We assembled a large dataset of fine-scale forest carbon stocks and multi-taxonomical diversity for temperate forest, where knowledge gaps are largest. We found little evidence that above-ground live carbon and species richness in temperate forests are congruent at the stand scale, which contrasts with most of the evidence from the tropics (Sullivan et al., 2017). For all taxa we investigated, we found that win-win species gradually replaced trade-off species with increasing above-ground live carbon levels, changes that aggregate biodiversity indices (e.g. richness) would fail to detect. In general, species and community-level change-points were neither congruent, nor equally abrupt across taxa, suggesting that leveraging co-benefits across taxonomic groups might be difficult. Overall, our results highlight that in temperate forests it may not be best to jointly pursue conservation and climate-change-mitigation goals at the stand scale. Rather, forest planners should establish local priorities to leverage potentially higher co-benefits at broader scales. Stand-scale priorities can be established by taking into account the taxon-specific carbon-biodiversity relationship and the share of win-win vs. trade-off species to establish a safe operating space to manipulate above-ground live carbon levels while avoiding undesired biodiversity loss.

Our work provides new insights into the shape, variability and context-specificity of the carbon-biodiversity relationship, especially for taxonomic groups that are rarely considered (e.g., fungi, lichens, saproxylic beetles and bryophytes). The correlation between carbon and biodiversity was overall relatively weak and highly variable across taxonomic groups. While previous research mostly reported a positive carbon-biodiversity relationship in tropical forest (Deere et al., 2018; Magnago et al., 2015; Sullivan et al., 2017), for temperate forests the evidence is still inconclusive and mainly based on tree species only (Potter & Woodall, 2014; Xian et al., 2015). The contrasting patterns observed across different taxonomic groups may explain the weak relationship between multidiversity and above-ground live carbon. On the one hand, carbon-dense, late-successional forests represent better habitat than open forests for many organisms. Birds, for instance, may benefit from high above-ground live carbon, especially in the presence of ecologically complex carbon, such as wide-branching canopies, large standing trees and stem cavities (Bouvet et al., 2016; Hatanaka et al., 2011). High above-ground live carbon also often correlates with high deadwood levels, which represents fundamental resources for wood-inhabiting fungi or beetles (Lassauce et al., 2011; Stokland et al., 2012). On the other hand, plant richness decreased with increasing above-ground live carbon stocks, likely as a consequence of the strong, asymmetrical competition for light exerted by few tree species on the herb-layer, which comprises the majority of plant species in temperate forests (Sabatini, Jiménez-Alfaro, Burrascano, & Blasi, 2014). Furthermore, in our assessment most variation occurred across forest sites, rather than along the above-ground live carbon gradient. This is in agreement with previous research, which highlighted the importance of broad-scale drivers, including macroclimate and the regional species pool, as determinants of forest fine-scale biodiversity (Jiménez-Alfaro et al., 2018; Sullivan et al., 2017).

Forest assemblage composition did change along the above-ground live carbon gradient, although species richness and multidiversity did not. With increasing above-ground live carbon, win-win species gradually replaced trade-off species, confirming that what constitutes suitable habitat conditions differs among species (Lindenmayer et al., 2005). Both in oak- and beech-dominated forests, the overall proportion of win-win species was similar to the proportion of trade-off species. This might concur at explaining why both multidiversity and scaled richness per taxa were relatively insensitive to increasing above-ground live carbon, since neither can discriminate between colonization and local extinction of species when these occur simultaneously and gradually in response to shifts in ecological conditions (Lindenmayer et al., 2005). As observed in tropical forests, the effect of carbon removal on species richness may be confounded by the increase in generalist species, so that a special focus on sensitive species, or species of conservation concern is recommended (Deere et al., 2018; Magnago et al., 2015). In our work, the conservation status of the vast majority (92%) of species was not available, especially for understudied taxa (i.e. lichens, fungi and bryophytes; IUCN, 2017). In the absence of species-level assessments, identifying win-win and trade-off species, provides a proxy for identifying sensitive taxa, and for calibrating biodiversity goals when managing temperate forests for climate-change mitigation.

Community-level change-points differed between win-win and trade-off species, as well as across taxa and forest types, suggesting that a clear ecological threshold along the carbon-stock gradient may not exist in temperate forests. When considering trade-off species, however, we found a relatively marked decrease between ~80-120 MgC/ha, and the change-points of different taxonomic groups were surprisingly similar across the two forest types. We interpret this result as the effect of canopy closure (i.e. the phase of forest succession when the canopies of individual trees overlap), which reduces light availability, buffers temperature variation, increases relative humidity, and nearly excludes wind at the forest floor (Franklin et al., 2002). These ecological changes may determine a shift in the species composition of the herb-layer and facilitate forest succession towards the dominance of shade tolerant species, while triggering bottom-up cascading effects on the whole trophic network (Kagata & Ohgushi, 2006). Furthermore, canopy closure kicks off self-thinning processes, which provide a first pulse of deadwood in early successional stands, thus favouring the colonization by saproxylic species (Lassauce et al., 2011; Stokland et al., 2012). This result suggests that a general pattern may exist in temperate forests and advises against assuming a linear positive carbon-biodiversity relationship in conservation actions (Di Marco et al. 2018).

We used a large, multi-taxonomic dataset collected across a broad geographical area and environmental gradient, which give us confidence on the generality of our result. Still, our analyses do not come without uncertainty. First, although comprehensive, our dataset suffers from a lack of detailed information on management and disturbance history, both of which influence biodiversity and carbon stock, and possibly their relationship (Paillet et al., 2010). Second, our results provide indication on the carbon-biodiversity relationship exclusively for natural or semi-natural forests, i.e. self-regenerated forests of native species. These forests are particularly relevant for biodiversity conservation, but in many European countries they are often replaced with forest plantations composed of very productive carbon-sinking species (e.g., *Picea abies, Pseudotsuga menziesii*, *Eucalyptus spp.*) that may show different carbon-biodiversity relationships (Pichancourt et al., 2014). Finally, our dataset is spatially nested in nature. Failing to treat data nestedness leads to the classical problem of pseudoreplication and increases the probability of type I error. To avoid this problem, we accounted for nestedness by stratifying the bootstrapping procedure when exploring change-points through TITAN. This approach, however, strongly reduces the power of the analysis, which means that our analysis probably detected only a conservative subset of the win-win and trade-off species.

Our work provides new understanding of how co-benefits between biodiversity and carbon storage might be leveraged in temperate forests. Three major implications for management derive from our results. First, biodiversity and above-ground live carbon cannot easily be simultaneously maximized in temperate forests at the scale of individual stands. Instead of seeking to integrate both goals at the stand scale (i.e., a land sharing strategy), co-benefits might be larger for strategies that seek to maximize co-benefits at broader scales by prioritizing either biodiversity conservation or carbon-storage goals at the stand scale (i.e., a land sparing strategy; Butsic & Kuemmerle, 2015; Edwards et al., 2014). Forest planners and managers should carefully evaluate whether to give priority to biodiversity conservation or other carbon-related goals, since maximizing forest carbon at the stand scale may only benefit some species, while harming others. Importantly, the scaling of trade-offs in temperate forests thus appears to differ fundamentally from tropical forests, at least those under management, where climate and biodiversity objectives may be effectively be integrated at the stand scale (Deere et al., 2018; Magnago et al., 2015; Ferreira et al, 2018).

Second, reconciling biodiversity and carbon objectives requires planning across scales. This includes assessments of which arrangement of management types delivers lowest trade-off or highest co-benefits (Law et al., 2017; Reside et al., 2017), while integrating stand-scale constraints (Pichancourt et al., 2014). Likewise, landscape-scale planning is needed to ensure heterogeneity in forest developmental stage and structure across stands (Schall et al., 2018), which should include set-asides (Bouget et al., 2014; Ferreira et al., 2018; Hatanaka et al., 2011), to ensure that trade-off assessments are robust over time. Encouraging the retention of blocks of undisturbed forest as a conservation priority within managed forests may represent an effective option for reconciling carbon-storage and conservation goals, while incorporating multiple environmental goals in forest management (Edwards et al., 2014). This would allow for the persistence of the full range of both win-win and trade-off species, and therefore maximizing multi-taxonomic diversity while optimizing carbon-stock allocation (Trentanovi, Campagnaro, Rizzi, & Sitzia, 2018). Third, rather than relying on synthetic indices of biodiversity, only an explicit consideration of all taxa of conservation concern will provide the full picture of how these taxa respond to the manipulation of forest structure and above-ground live carbon stocks in temperate forests (Villard & Jonsson, 2009).

### Acknowledgments:

F.M.S. was funded by the European Union's Horizon 2020 research and innovation programme under the *Marie Skłodowska-Curie*, grant agreement No. 658876. Sapienza University of Rome provided the funding for the networking activities that allowed this joint research (Ricerche Universitarie 2015 - prot. C26A15KE2T). Data collected in central and southern Italy derive from the LIFE+ project FAGUS (11/NAT/IT/00135), those relative to Hungary from Hungarian Research Fund (OTKA K79158), Őrség

National Park Directorate; for the French Alps funding came from IRSTEA and from the Conseil Général de l'Isère and Bauges Natural Regional Park; data for the rest of France were collected by funding of the French Ministry in charge of Ecology (Convention Cemagref-DEB (MEEDDAT), Action GNB and the program "Biodiversité, Gestion Forestière et Politiques Publiques" (BGF), convention GNB 10-MBGD-BGF-1-CVS-092, no CHORUS 2100 214 651) and the Office National des Forêts (Convention ONF-Cemagref, Action 5, 2008); data for northern Italy was collected with funding provided by the Italian Ministry of Agricultural, Food and Forestry Policies, State Forestry Corps (Project Managers: A. Andrighetti and D. Campedel), research agreement No. 767/ 2008 (TS, Principal Investigator). Above all we wish to thank all the people that collaborated in the fieldwork and in the species identification and that solved nomenclature issues:  G. Antonini, M.M. Azzella, H. Brustel, M. Cassol, O. Courtin, M. Dal Cortivo, J. Delnatte, L. Facioni, G. Favier, E. Gatti, G. Gobbo, J. Haran, I. Király, G. Kutszegi, S. Labonne, F. Lakatos, E. Lattanzi, F. Lebagousse, D. Lunghini, Z. Mag, O. Maggi, S. Márialigeti, C. Moliard, B. Németh, T. Noblecourt, B. Nusillard, F. Padovan, F. Parisi, S. Ravera, O. Rose, I. Siller, M. Sommacal, P. Tardif, A. Tilia, F. Tinya, M.Varaschin.

## References

Allan, E., Bossdorf, O., Dormann, C. F., Prati, D., Gossner, M. M., Tscharntke, T., . . . Boch, S. (2014). Interannual variation in land-use intensity enhances grassland multidiversity. *Proceedings of the National Academy of Sciences, 111*(1), 308-313.

Baker, M. E., & King, R. S. (2010). A new method for detecting and interpreting biodiversity and ecological community thresholds. *Methods in Ecology and Evolution, 1*(1), 25-37.

Beaudrot, L., Kroetz, K., Alvarez‐Loayza, P., Amaral, I., Breuer, T., Fletcher, C., . . . Marshall, A. R. (2016). Limited carbon and biodiversity co‐benefits for tropical forest mammals and birds. *Ecological Applications, 26*(4), 1098-1111.

Bouget, C., Parmain, G., Gilg, O., Noblecourt, T., Nusillard, B., Paillet, Y., . . . Gosselin, F. (2014). Does a set‐aside conservation strategy help the restoration of old‐growth forest attributes and recolonization by saproxylic beetles? *Animal Conservation, 17*(4), 342-353.

Bouvet, A., Paillet, Y., Archaux, F., Tillon, L., Denis, P., Gilg, O., & Gosselin, F. (2016). Effects of forest structure, management and landscape on bird and bat communities. *Environmental Conservation, 43*(2), 148-160.

Boysen, L. R., Lucht, W., & Gerten, D. (2017). Trade‐offs for food production, nature conservation and climate limit the terrestrial carbon dioxide removal potential. *Global Change Biology, 23*(10), 4303-4317.

Bremer, L. L., & Farley, K. A. (2010). Does plantation forestry restore biodiversity or create green deserts? A synthesis of the effects of land-use transitions on plant species richness. *Biodiversity and Conservation, 19*(14), 3893-3915.

Burrascano, S., Chytrý, M., Kuemmerle, T., Giarrizzo, E., Luyssaert, S., Sabatini, F. M., & Blasi, C. (2016). Current European policies are unlikely to jointly foster carbon sequestration and protect biodiversity. *Biological Conservation, 201*, 370-376. doi:http://dx.doi.org/10.1016/j.biocon.2016.08.005

Burrascano, S., De Andrade, R. B., Paillet, Y., Ódor, P., Antonini, G., Bouget, C., . . . Blasi, C. (2018). Congruence across taxa and spatial scales: Are we asking too much of species data? *Global Ecology and Biogeography*. doi:10.1111/geb.12766

Bustamante, M. M., Roitman, I., Aide, T. M., Alencar, A., Anderson, L. O., Aragão, L., . . . Chambers, J. (2016). Toward an integrated monitoring framework to assess the effects of tropical forest degradation and recovery on carbon stocks and biodiversity. *Global Change Biology, 22*(1), 92-109.

Butsic, V., & Kuemmerle, T. (2015). Using optimization methods to align food production and biodiversity conservation beyond land sharing and land sparing. *Ecological Applications*, 25(3), 589-595.

Cavanaugh, K. C., Gosnell, J. S., Davis, S. L., Ahumada, J., Boundja, P., Clark, D. B., . . . Andelman, S. (2014). Carbon storage in tropical forests correlates with taxonomic diversity and functional dominance on a global scale. *Global Ecology and Biogeography, 23*(5), 563-573. doi:10.1111/geb.12143

CBD (2006). *Framework for monitoring implementation of the achievement of the 2010 target and integration of targets into the thematic programmes of work*: (Decision VIII/15, COP 8, 2006); available at www.cbd.int/decisions/.

Chamberlain, S. (2017). rredlist: 'IUCN' Red List Client. R package version 0.4.0. https://CRAN.R-project.org/package=rredlist.

Colwell, R. K., Chao, A., Gotelli, N. J., Lin, S.-Y., Mao, C. X., Chazdon, R. L., & Longino, J. T. (2012). Models and estimators linking individual-based and sample-based rarefaction, extrapolation and comparison of assemblages. *Journal of plant ecology, 5*(1), 3-21.

Deere, N. J., Guillera‐Arroita, G., Baking, E. L., Bernard, H., Pfeifer, M., Reynolds, G., . . . Struebig, M. J. (2018). High Carbon Stock forests provide co‐benefits for tropical biodiversity. *Journal of Applied Ecology, 55*(2), 997-1008.

Di Marco, M., Watson, J. E., Currie, D. J., Possingham, H. P., & Venter, O. (2018). The extent and predictability of the biodiversity–carbon correlation. *Ecology Letters*.

Dufrêne, M., & Legendre, P. (1997). Species assemblages and indicator species: the need for a flexible asymmetrical approach. *Ecological Monographs, 61*, 345-366.

Edwards, D. P., Gilroy, J. J., Woodcock, P., Edwards, F. A., Larsen, T. H., Andrews, D. J. R., . . . Wilcove, D. S. (2014). Land-sharing versus land-sparing logging: reconciling timber extraction with biodiversity conservation. *Global Change Biology, 20*(1), 183-191. doi:doi:10.1111/gcb.12353

EEA (2006). *European forest types. Categories and types for sustainable forest management reporting and policy*. Copenhagen, Denmark. Retrieved from: https://www.eea.europa.eu/publications/technical_report_2006_9

Elith, J., Leathwick, J. R., & Hastie, T. (2008). A working guide to boosted regression trees. *Journal of Animal Ecology, 77*(4), 802-813.

European Commission. (2013). *A new EU Forest Strategy: for forests and the forest-based sector*. Brussels: European Commission, COM(2013) 659 Final.

Evans, P. M., Newton, A. C., Cantarello, E., Martin, P., Sanderson, N., Jones, D. L., . . . Fuller, L. (2017). Thresholds of biodiversity and ecosystem function in a forest ecosystem undergoing dieback. *Scientific Reports, 7*(1), 6775. doi:10.1038/s41598-017-06082-6

FAO. (2015). *Global Forest Resources Assessment 2015. Desk reference*. Rome, Italy.

Federici, S., Vitullo, M., Tulipano, S., De Lauretis, R., & Seufert, G. (2008). An approach to estimate carbon stocks change in forest carbon pools under the UNFCCC: the Italian case. *Iforest-Biogeosciences and Forestry, 1*(2), 86-95.

Ferreira, J., Lennox, G. D., Gardner, T. A., Thomson, J. R., Berenguer, E., Lees, A. C., . . . Barlow, J. (2018). Carbon-focused conservation may fail to protect the most biodiverse tropical forests. *Nature Climate Change*. doi:10.1038/s41558-018-0225-7

Fick, S. E., & Hijmans, R. J. (2017). WorldClim 2: new 1‐km spatial resolution climate surfaces for global land areas. *International Journal of Climatology*.

FOREST EUROPE (2015). *State of Europe's Forests 2015*. Madrid, Spain. Retrieved from: https://www.foresteurope.org/docs/fullsoef2015.pdf

Franklin, J. F., Spies, T. A., Van Pelt, R., Carey, A. B., Thornburgh, D. A., Berg, D. R., . . . Chen, J. Q. (2002). Disturbances and structural development of natural forest ecosystems with silvicultural implications, using Douglas-Fir forests as an example. *Forest Ecology and Management, 155*(1-3), 399-423.

Gardner, T. A., Burgess, N. D., Aguilar-Amuchastegui, N., Barlow, J., Berenguer, E., Clements, T., . . . Vieira, I. C. G. (2012). A framework for integrating biodiversity concerns into national REDD+ programmes. *Biological Conservation, 154*, 61-71. doi:https://doi.org/10.1016/j.biocon.2011.11.018

Hatanaka, N., Wright, W., Loyn, R. H., & Mac Nally, R. (2011). 'Ecologically complex carbon' - linking biodiversity values, carbon storage and habitat structure in some austral temperate forests. *Global Ecology and Biogeography, 20*(2), 260-271. doi:10.1111/j.1466-8238.2010.00591.x

Hijmans, R. J., Phillips, S., Leathwick, J., & Elith, J. (2011). Package 'dismo'. Available online at: http://cran.r-project.org/web/packages/dismo/index.html.

Hsieh, T., Ma, K., & Chao, A. (2016). iNEXT: an R package for rarefaction and extrapolation of species diversity (Hill numbers). *Methods in Ecology and Evolution, 7*(12), 1451-1456.

IUCN (2017). The IUCN Red List of Threatened Species. Version 2017-3. <http://www.iucnredlist.org>. Downloaded on 05 December 2017.

Jiménez-Alfaro, B., Girardello, M., Chytrý, M., Svenning, J.-C., Willner, W., Gégout, J.-C., . . . Wohlgemuth, T. (2018). History and environment shape species pools and community diversity in European beech forests. *Nature Ecology & Evolution, 2*(3), 483-490. doi:10.1038/s41559-017-0462-6

Kagata, H., & Ohgushi, T. (2006). Bottom-up trophic cascades and material transfer in terrestrial food webs. *Ecological Research, 21*(1), 26-34.

Larrieu, L., Gosselin, F., Archaux, F., Chevalier, R., Corriol, G., Dauffy-Richard, E., . . . Savoie, J.-M. (2018). Cost-efficiency of cross-taxon surrogates in temperate forests. *Ecological Indicators, 87*, 56-65.

Lassauce, A., Paillet, Y., Jactel, H., & Bouget, C. (2011). Deadwood as a surrogate for forest biodiversity: Meta-analysis of correlations between deadwood volume and species richness of saproxylic organisms. *Ecological Indicators, 11*(5), 1027-1039. doi:10.1016/j.ecolind.2011.02.004

Law, E. A., Bryan, B. A., Meijaard, E., Mallawaarachchi, T., Struebig, M. J., Watts, M. E., & Wilson, K. A. (2017). Mixed policies give more options in multifunctional tropical forest landscapes. *Journal of Applied Ecology, 54*(1), 51-60.

Li, W., Xu, F., Zheng, S., Taube, F., & Bai, Y. (2017). Patterns and thresholds of grazing‐induced changes in community structure and ecosystem functioning: species‐level responses and the critical role of species traits. *Journal of Applied Ecology, 54*(3), 963-975.

Lindenmayer, D. B., Fischer, J., & Cunningham, R. B. (2005). Native vegetation cover thresholds associated with species responses. *Biological Conservation, 124*(3), 311-316. doi:https://doi.org/10.1016/j.biocon.2005.01.038

Magnago, L. F. S., Magrach, A., Laurance, W. F., Martins, S. V., Meira‐Neto, J. A. A., Simonelli, M., & Edwards, D. P. (2015). Would protecting tropical forest fragments provide carbon and biodiversity cobenefits under REDD+? *Global Change Biology, 21*(9), 3455-3468.

MEA (2005). *Millennium Ecosystem Assessment. Ecosystems and Human Well-being: Synthesis*. Washington, DC: Island Press.

Mori, A. S., Lertzman, K. P., & Gustafsson, L. (2017). Biodiversity and ecosystem services in forest ecosystems: a research agenda for applied forest ecology. *Journal of Applied Ecology, 54*(1), 12-27. doi:10.1111/1365-2664.12669

NASA (2006). *Shuttle Radar Topography Mission. <http://www.jpl.nasa.gov/srtm> (accessed 01.09.16)*.

Paillet, Y., Berges, L., Hjalten, J., Odor, P., Avon, C., Bernhardt-Romermann, M., . . . Virtanen, R. (2010). Biodiversity Differences between Managed and Unmanaged Forests: Meta-Analysis of Species Richness in Europe. *Conservation Biology, 24*(1), 101-112.

Pan, Y. D., Birdsey, R. A., Fang, J. Y., Houghton, R., Kauppi, P. E., Kurz, W. A., . . . Hayes, D. (2011). A Large and Persistent Carbon Sink in the World's Forests. *Science, 333*(6045), 988-993. doi:10.1126/science.1201609

Panagos, P., Jones, A., Bosco, C., & Kumar, P. S. (2011). European digital archive on soil maps (EuDASM): preserving important soil data for public free access. *International Journal of Digital Earth, 4*(5), 434-443.

Pellegrini, A. F. A., Socolar, J. B., Elsen, P. R., & Giam, X. (2016). Trade‐offs between savanna woody plant diversity and carbon storage in the Brazilian Cerrado. *Global Change Biology, 22*(10), 3373-3382. doi:doi:10.1111/gcb.13259

Pichancourt, J. B., Firn, J., Chadès, I., & Martin, T. G. (2014). Growing biodiverse carbon-rich forests. *Global Change Biology, 20*(2), 382-393. doi:10.1111/gcb.12345

Potter, K. M., & Woodall, C. W. (2014). Does biodiversity make a difference? Relationships between species richness, evolutionary diversity, and aboveground live tree biomass across U.S. forests. *Forest Ecology and Management, 321*, 117-129. doi:https://doi.org/10.1016/j.foreco.2013.06.026

Ratcliffe, S., Wirth, C., Jucker, T., der Plas, F., Scherer‐Lorenzen, M., Verheyen, K., . . . Ohse, B. (2017). Biodiversity and ecosystem functioning relations in European forests depend on environmental context. *Ecology Letters*.

Reside, A. E., VanDerWal, J., & Moran, C. (2017). Trade-offs in carbon storage and biodiversity conservation under climate change reveal risk to endemic species. *Biological Conservation, 207*, 9-16.

Sabatini, F. M., Burrascano, S., Azzella, M. M., Barbati, A., De Paulis, S., Di Santo, D., . . . Blasi, C. (2016). One taxon does not fit all: Herb-layer diversity and stand structural complexity are weak predictors of biodiversity in Fagus sylvatica forests. *Ecological Indicators, 69*, 126-137. doi:http://dx.doi.org/10.1016/j.ecolind.2016.04.012

Sabatini, F. M., Jiménez-Alfaro, B., Burrascano, S., & Blasi, C. (2014). Drivers of herb-layer species diversity in two unmanaged temperate forests in northern Spain. *Community Ecology, 15*(2), 147-157. doi:10.1556/ComEc.15.2014.2.3

Sasaki, T., Furukawa, T., Iwasaki, Y., Seto, M., & Mori, A. S. (2015). Perspectives for ecosystem management based on ecosystem resilience and ecological thresholds against multiple and stochastic disturbances. *Ecological Indicators, 57*, 395-408.

Schall, P., Gossner, M. M., Heinrichs, S., Fischer, M., Boch, S., Prati, D., . . . Böhm, S. (2018). The impact of even‐aged and uneven‐aged forest management on regional biodiversity of multiple taxa in European beech forests. *Journal of Applied Ecology*.

Sollmann, R., Mohamed, A., Niedballa, J., Bender, J., Ambu, L., Lagan, P., . . . Gardner, B. (2017). Quantifying mammal biodiversity co‐benefits in certified tropical forests. *Diversity and Distributions, 23*(3), 317-328.

Stokland, J. N., Siitonen, J., & Jonsson, B. G. (2012). *Biodiversity in dead wood*: Cambridge University Press.

Strassburg, B. B. N., Kelly, A., Balmford, A., Davies, R. G., Gibbs, H. K., Lovett, A., . . . Rodrigues, A. S. L. (2010). Global congruence of carbon storage and biodiversity in terrestrial ecosystems. *Conservation Letters, 3*(2), 98-105. doi:10.1111/j.1755-263X.2009.00092.x

Sullivan, M. J. P., Talbot, J., Lewis, S. L., Phillips, O. L., Qie, L., Begne, S. K., . . . Zemagho, L. (2017). Diversity and carbon storage across the tropical forest biome. *Scientific Reports, 7*, 39102. doi:10.1038/srep39102

Trentanovi, G., Campagnaro, T., Rizzi, A., & Sitzia, T. (2017). Synergies of planning for forests and planning for Natura 2000: Evidences and prospects from northern Italy. *Journal for Nature Conservation, 43*, 239-249. doi:10.1016/j.jnc.2017.07.006

Villard, M.-A., & Jonsson, B. G. (2009). Tolerance of focal species to forest management intensity as a guide in the development of conservation targets. *Forest Ecology and Management, 258*, S142-S145. doi:https://doi.org/10.1016/j.foreco.2009.08.034

Xian, W., Xiangping, W., Zhiyao, T., Zehao, S., Chengyang, Z., Xinli, X., & Jingyun, F. (2015). The relationship between species richness and biomass changes from boreal to subtropical forests in China. *Ecography, 38*(6), 602-613. doi:doi:10.1111/ecog.00940

Zilliox, C., & Gosselin, F. (2014). Tree species diversity and abundance as indicators of understory diversity in French mountain forests: variations of the relationship in geographical and ecological space. *Forest Ecology and Management, 321*, 105-116.

## Supplementary material

Table S1 – Description of study sites.

Table S2 – Main characteristics of the sampling protocols across different datasets.

Table S3 – Diagnostics of boosted regression tree models modelling multidiversity and the scaled richness of different taxonomic groups.

Figure S1 – Partial dependency plots of the relationship between the scaled richness of beetles and the six top-performing explanatory variables as modelled through a Boosted Regression Tree.

Figure S2 – Partial dependency plots for birds.

Figure S3 – Partial dependency plots for bryophytes.

Figure S4 - Partial dependency plots for fungi.

Figure S5 - Partial dependency plots for lichens.

Figure S6 - Partial dependency plots for plants.

Figure S7 – Partial dependency plots for multidiversity.

Figure S8 – Diving board plot of species-specific change-points for pure and reliable indicator species in oak-dominated forests.

Figure S9 – Diving board plot of species-specific change-points for pure and reliable indicator species in beech forests.

Table S1 – Description of study sites.

| Stand Name | Substrate | Number of Plots | Annual Mean T °C | Annual rainfall mm | Elevation m a.s.l. | Aspect ° | Slope ° | Latitude | Longitude |
|---|---|---|---|---|---|---|---|---|---|
| Ventron | Limestone | 22 | 9.0 | 800-900 | 455 | 229 | 5 | 47.818 | 6.798 |
| Auberive | Limestone | 14 | 6.9 | 1300-1400 | 1043 | 189 | 16 | 47.583 | 3.649 |
| Ballons-Comtois | Granit | 9 | 10.9 | 650-750 | 195 | 222 | 4 | 46.235 | 12.237 |
| Bois du Parc | Limestone | 9 | 7.2 | 1300-1500 | 1255 | 141 | 11 | 46.146 | -0.386 |
| Cajada forest - Dolomiti Bellunesi National Park | Dolomite/ Limestone | 18 | 12.0 | 800-900 | 89 | 207 | 2 | 47.090 | 5.050 |
| Chizé | Limestone | 10 | 10.9 | 700-800 | 231 | 205 | 2 | 47.226 | 4.939 |
| Citeaux | Alluvial deposits | 8 | 9.1 | 800-900 | 572 | 197 | 12 | 48.421 | 2.660 |
| Combe Lavaux | Limestone | 25 | 10.5 | 600-700 | 138 | 172 | 3 | 45.145 | 5.506 |
| Fontainebleau | Acidic sands | 18 | 7.0 | 1200-1500 | 1232 | 189 | 15 | 45.339 | 5.791 |
| Vercors | Limestone | 31 | 6.8 | 1300-1900 | 1136 | 245 | 19 | 45.701 | 6.167 |
| Chartreuse | Limestone | 29 | 7.2 | 1350-1850 | 1107 | 266 | 22 | 46.299 | 5.986 |
| Bauges | Limestone | 6 | 7.5 | 1300-1400 | 1101 | 109 | 25 | 48.104 | 4.188 |
| Haute-Chaine du Jura | Limestone | 12 | 10.0 | 650-750 | 175 | 170 | 2 | 46.892 | 16.308 |
| Haut Tuileau | Alluvial deposits | 25 | 9.2 | 700-800 | 308 | 134 | 5 | 40.465 | 15.339 |
| Orseg | gravel/loess | 24 | 10.1 | 718-1250 | 1272 | 177 | 17 | 40.257 | 15.344 |
| Cilento N | Limestone | 12 | 11.2 | 700 | 1246 | 175 | 17 | 42.508 | 13.514 |
| Cilento S | Limestone/ Flysch | 19 | 9.1 | 1062-1097 | 1354 | 263 | 17 | 44.123 | 5.820 |
| Gran Sasso | Limestone | 7 | 8.0 | 1000-1100 | 1448 | 109 | 31 | 48.671 | 1.763 |
| Lure | Limestone | 16 | 10.0 | 600-700 | 164 | 180 | 2 | 46.309 | 12.302 |
| Rambouillet | Acidic sands | 11 | 7.2 | 1300-1500 | 1236 | 306 | 19 | 47.939 | 6.931 |
| Val Tovanella Nature Reserve | Dolomite/ Limestone | 3 | 8.9 | 1200-1300 | 956 | 266 | 22 | 48.757 | 2.250 |
| Verrières | Acidic sands marls | 7 | 9.6 | 600-700 | 177 | 203 | 2 | 44.181 | 5.261 |
| Ventoux | Limestone | 9 | 7.9 | 1000-1100 | 1356 | 187 | 25 | 47.818 | 6.798 |

*Table S2 - Main characteristics of the sampling protocols across different datasets.*

| | Dataset | IT_Cilento | IT_FAGUS | IT_NE | HU | FR_YP | FR_Alps |
|---|---|---|---|---|---|---|---|
| | Country | Italy | Italy | Italy | Hungary | France | France |
| | Responsible | SB | SB | TS | PÓ | YP | PJ |
| **Taxon** | **Detail** | | | | | | |
| **Plants** | Sampling unit size shape | 1256 m² circular plot | 1256 m² circular plot | 490 m² circular plot | 900 m² square plot | 1000 m² circular plot | 314 m² circular plot |
| | Season/year | Summer 2007 | Summer 2013 | Summer 2007 | Summer 2006 | Spring-summer 2008-2013 | Spring-Summer 2014 |
| | Reference | Burrascano et al. (2011) | Sabatini et al. (2016) | Sitzia et al. (2012) | Márialigeti et al. (2016) | | Janssen et al. (2017) |
| **Bryophytes** | Sampling unit size shape | four 154 m² circular plots | | | 900 m² square plot | 1256 m² circular plot | |
| | Substrates | wood, ground | | | wood, ground | wood, ground | |
| | Season/year | Summer 2008 | | | Summer 2006-2009 | Spring and Autumn 2008-2013 | |
| | Reference | Blasi et al. (2010); Brunialti et al. (2010) | | | Király et al. (2013); Márialigeti et al. (2009); Ódor et al. (2013) | | |
| **Lichens** | Sampling unit size shape | three 154 m² circular plots | 1256 m² circular plot | 490 m² circular plot | 900 m² square plot | | 1256 m² square plot |
| | Substrates | 3 trees (DBH > 16cm) in each subplot | 3 trees (DBH > 16cm) in each subplot | all trees | all trees (DBH > 20cm) | | 6 largest trees |
| | Season/year | Summer 2008 | Summer 2013 | Summer 2007 | Summer 2010 | | Summer 2015 |
| | Reference | Blasi et al. (2010); Brunialti et al. (2010) | Sabatini et al. (2016) | Nascimbene et al. (2013); Sitzia et al. (2017) | Király et al. (2013); Ódor et al. (2013) | | |
| **Birds** | Sampling unit size shape | 1256 m² circular plot | 1256 m² circular plot | 490 m² circular plot | 31,400 m² circular plot | 31,400 m² circular plot | |
| | Season/year | Summer 2008 | Summer 2013 | Summer 2010-2011 | Spring-Summer 2006 | Spring-summer 2008-2013 | |
| | Methodology | 20-min pointcount | 10-min pointcount | 10-min pointcount | pointcount | 5-min pointcount | |
| | Frequency of sampling: | 15 days | once | twice a year | twice a year | twice a year | |

|  | Reference | Blasi et al. (2010) | Sabatini et al. (2016) | Sitzia et al. (2017) | Mag & Ódor (2015) | Bouvet et al. (2016) | |
|---|---|---|---|---|---|---|---|
| **Beetles** | Sampling unit size shape | 1256 m² circular plot | 1256 m² circular plot | 490 m² circular plot | 900 m² square plot | 1256 m² circular plot | 1256 m² square plot |
|  | Season/year | Spring and Summer 2008 | Summer 2013 | Spring-Summer-Autumn 2009 | Spring-Summer 2010 | Spring-Summer 2008 | Spring-Summer 2014 |
|  | Methodology | windowflight traps | window flight traps, emerging traps | windowflight traps | emerging traps | windowflight traps | windowflight traps, Winkler extractors |
|  | Frequency of sampling | monthly | monthly | twice a month | only once | monthly (for 3 months) | monthly |
|  | Reference | Blasi et al. (2010); Persiani et al. (2010) | Sabatini et al. (2016) | Sitzia et al. (2015; 2017) |  | Bouget et al. (2016) | Janssen et al. (2016) |
| **Fungi** | Sampling unit size shape | four 154 m² circular plots | 530 m² circular plot | 490 m² circular plot | 900 m² square plot | 1256 m² circular plot |  |
|  | Substrates | wood > 10 cm diameter | wood > 10 cm diameter | wood | wood > 10 cm diameter | wood |  |
|  | Season/year | Summer 2008 | Autumn 2013 | Summer-Autumn 2009-2010 | Spring-Autumn 2010-Summer 2009 | Spring-Autumn 2008-2013 |  |
|  | Reference | Blasi et al. (2010); Persiani et al. (2010) | Sabatini et al. (2016) | Sitzia et al. (2017) | Kutszegi et al. (2015) |  |  |
| **Forest structure** | Plot size for living trees (DBH>10 cm) | concentric circular areas with a radius of 13 and 20 m | concentric circular areas with a radius of 13 and 20 m | 491 m² | 40 x 40 m plot | Combined fixed surface (314 m²) and fixed angle (2%) | 10-m-radius subplot for DBH > 10 cm; 20-m-radius subplot for DBH > 30 cm; |
|  | Plot size for deadwood (mid diameter >10 cm | Circular areas with a radius of 13 m | Circular areas with a radius of 13 m | logs: 491 m²; stumps/snags 736 m² | 40 x 40 m plot | Combined line intersect sampling (3 X 10m) and fixed surface (1256 m²) | 10-m-radius subplot for DBH > 10 cm; 20-m-radius subplot for DBH > 30 cm; |
|  | Trees with height measurement | 20 height samples per main species per plot | 20 height samples per main species per plot | All | All | 5 largest trees (Dominant height) | No height measurement taken |
|  | Allometric equations | Tabacchi et al. (2011) | Tabacchi et al. (2011) | Castellani et al. (1984) | Sopp & Kolozs (2000) | Paillet et al. (2015) | In mountain forests, ALGAN table: softwood=11; hardwood=8 |

*Additional references cited in Table S2:*

Blasi C, Marchetti M, Chiavetta U *et al.* (2010) Multi-taxon and forest structure sampling for identification of indicators and monitoring of old-growth forest. Plant Biosystems, **144**, 160-170.

Bouvet A, Paillet Y, Archaux F, Tillon L, Denis P, Gilg O, Gosselin F (2016) Effects of forest structure, management and landscape on bird and bat communities. Environmental Conservation, **43**, 148-160.

Brunialti G, Frati L, Aleffi M, Marignani M, Rosati L, Burrascano S, Ravera S (2010) Lichens and bryophytes as indicators of old-growth features in Mediterranean forests. Plant Biosystems, **144**, 221-233.

Burrascano S, Sabatini FM, Blasi C (2011) Testing indicators of sustainable forest management on understorey composition and diversity in southern Italy through variation partitioning. Plant Ecology, **212**, 829-841.

Castellani C, Scrinzi G, Tabacchi G, Tosi V (1984) Inventario Forestale Nazionale Italiano (IFNI) Tavole di cubatura a doppia entrata. Istituto Sperimentale per l'Assestamento Forestale e per l'Alpicoltura, Trento.

Janssen, P., S. Bec, M. Fuhr, P. Taberlet, J. J. Brun, and C. Bouget. 2018. Present conditions may mediate the legacy effect of past land-use changes on species richness and composition of above- and below-ground assemblages. Journal of Ecology 106:306-318.

Janssen, P., E. Cateau, M. Fuhr, B. Nusillard, H. Brustel, and C. Bouget. 2016. Are biodiversity patterns of saproxylic beetles shaped by habitat limitation or dispersal limitation? A case study in unfragmented montane forests. Biodiversity and Conservation 25:1167-1185.

Kiraly I, Nascimbene J, Tinya F, Odor P (2013) Factors influencing epiphytic bryophyte and lichen species richness at different spatial scales in managed temperate forests. Biodiversity and Conservation, **22**, 209-223.

Kutszegi G, Siller I, Dima B *et al.* (2015) Drivers of macrofungal species composition in temperate forests, West Hungary: functional groups compared. Fungal Ecology, **17**, 69-83.

Mag Z, Ódor P (2015) The effect of stand-level habitat characteristics on breeding bird assemblages in Hungarian temperate mixed forests. Community Ecology, **16**, 156-166.

Márialigeti S, Németh B, Tinya F, Ódor P (2009) The effects of stand structure on ground-floor bryophyte assemblages in temperate mixed forests. Biodiversity and Conservation, **18**, 2223.

Márialigeti S, Tinya F, Bidló A, Ódor P (2016) Environmental drivers of the composition and diversity of the herb layer in mixed temperate forests in Hungary. Plant Ecology, **217**, 549-563.

Nascimbene J, Dainese M, Sitzia T (2013) Contrasting responses of epiphytic and dead wood-dwelling lichen diversity to forest management abandonment in silver fir mature woodlands. Forest Ecology and Management, **289**, 325-332.

Ódor P, Király I, Tinya F, Bortignon F, Nascimbene J (2013) Patterns and drivers of species composition of epiphytic bryophytes and lichens in managed temperate forests. Forest Ecology and Management, **306**, 256-265.

Persiani AM, Audisio P, Lunghini D et al. (2010) Linking taxonomical and functional biodiversity of saproxylic fungi and beetles in broad-leaved forests in southern Italy with varying management histories. Plant Biosystems, **144**, 250-261.

Sabatini FM, Burrascano S, Azzella MM et al. (2016) One taxon does not fit all: Herb-layer diversity and stand structural complexity are weak predictors of biodiversity in Fagus sylvatica forests. Ecological Indicators, **69**, 126-137.

Sitzia T, Trentanovi G, Dainese M, Gobbo G, Lingua E, Sommacal M (2012) Stand structure and plant species diversity in managed and abandoned silver fir mature woodlands. Forest Ecology and Management, **270**, 232-238.

Sitzia T, Campagnaro T, Dainese M et al. (2017) Contrasting multi-taxa diversity patterns between abandoned and non-intensively managed forests in the southern Dolomites. Iforest-Biogeosciences and Forestry, **10**, 845.

Sitzia T, Campagnaro T, Gatti E, Sommacal M, Kotze D (2015) Wildlife conservation through forestry abandonment: responses of beetle communities to habitat change in the Eastern Alps. European Journal of Forest Research, **134**, 511-524.

Sopp L, Kolozs L (2000) Fatömegszámítási táblázatok [Tables for calculating wood volume] Budapest, Állami Erdészeti Szolgálat.

Tabacchi G, Di Cosmo L, Gasparini P, Morelli S (2011) Stima del volume e della fitomassa delle principali specie forestali italiane. Equazioni di previsione, tavole del volume e tavole della fitomassa arborea epigea. Consiglio per la Ricerca e la sperimentazione in Agricoltura, Unità di Ricerca per il Monitoraggio e la Pianificazione Forestale. Trento. 412 pp. Trento: Consiglio per la Ricerca e la sperimentazione in Agricoltura, Unita di Ricerca per il Monitoraggio e la Pianificazione Forestale (in Italian).

***Nomenclatures and checklists***

Animals: 1) http://www.fauna-eu.org/ 2) http://www.gbif.org/ 3) http://www.organismnames.com/ 4) https://inpn.mnhn.fr/accueil/index

Bryophytes: Hodgetts N (2015) Checklist and Country Status of European Bryophytes: Towards a New Red List for Europe, National Parks and Wildlife Service.

Fungi: http://www.indexfungorum.org/names/names.asp

Lichens: Nimis, P.L., Martellos, S., 2008. ITALIC – The Information System on Italian Lichens,Version 4.0.  University of Trieste, Dept. of Biology, IN4.0/1(http://dbiodbs.univ.trieste.it/http://dbiodbs.univ.trieste.it/).

Plants: The Plant List (2013). Version 1.1. Published on the Internet; http://www.theplantlist.org/ (accessed July 2016).

*Table S3 – Diagnostics of boosted regression tree models modelling multidiversity and the scaled richness of different taxonomic groups.*

|  | Number of observations | Learning rate | Number of trees | Total Deviation | Residual Deviation | Cross-Validated correlation (mean) | Cross-Validated correlation (standard error) |
|---|---|---|---|---|---|---|---|
| **Multidiversity** | 352 | 0.0025 | 1450 | 0.005 | 0.001 | 0.786 | 0.025 |
| **Beetles** | 307 | 0.005 | 650 | 0.010 | 0.002 | 0.785 | 0.020 |
| **Birds** | 272 | 0.0025 | 1450 | 0.024 | 0.004 | 0.838 | 0.021 |
| **Bryophytes** | 180 | 0.005 | 800 | 0.011 | 0.003 | 0.722 | 0.040 |
| **Fungi** | 248 | 0.05 | 5550 | 0.007 | 0.000 | 0.711 | 0.045 |
| **Lichens** | 179 | 0.01 | 200 | 0.011 | 0.006 | 0.528 | 0.067 |
| **Plants** | 352 | 0.01 | 250 | 0.009 | 0.004 | 0.668 | 0.039 |

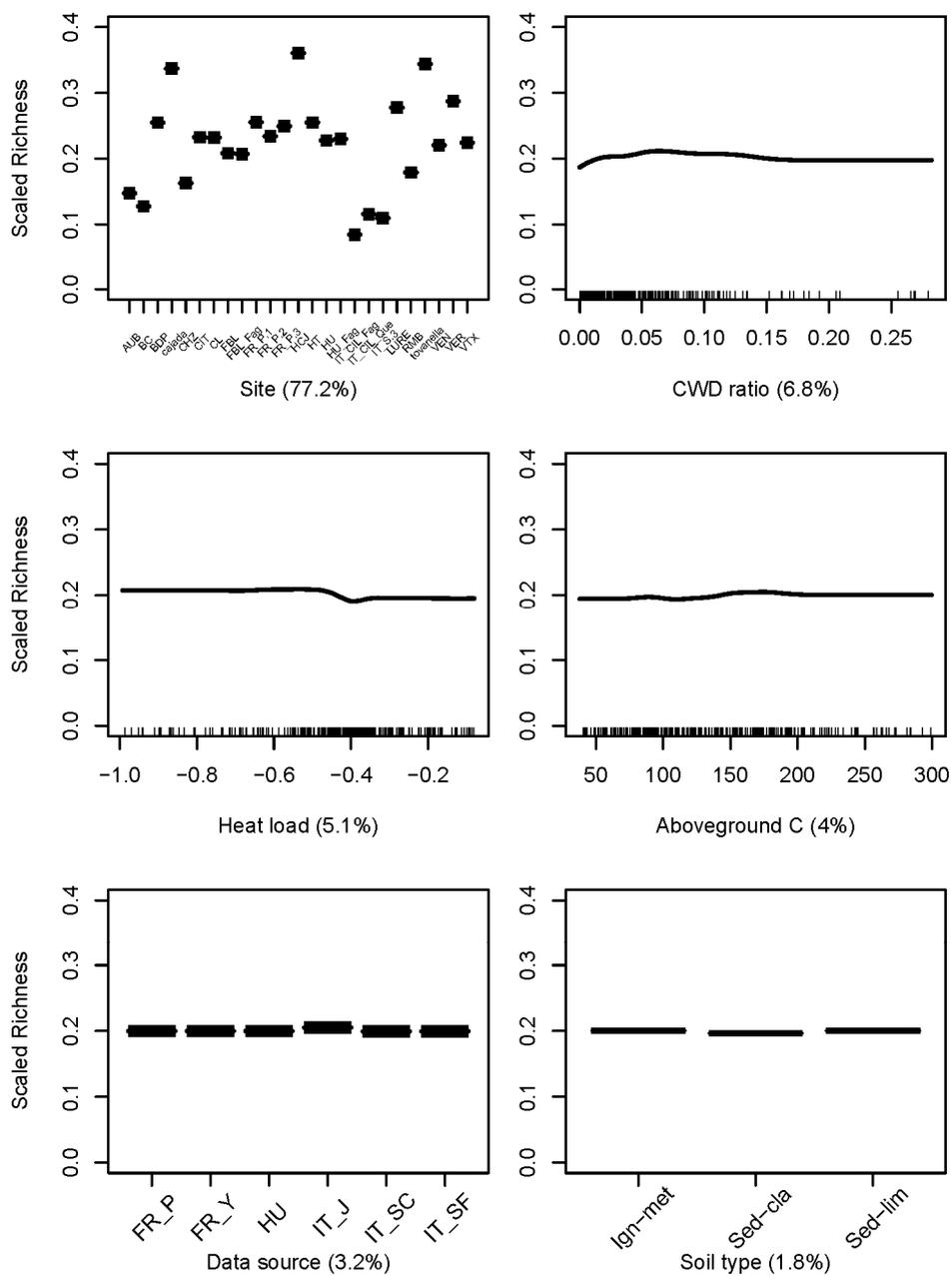

*Figure S1 – Partial dependency plots of the relationship between the scaled richness of beetles and the six top-performing explanatory variables as modelled through a Boosted Regression Tree. Scaled richness represents the fraction of species of the estimated species pool size observed in a given plot. Tick marks on the x-axis represent above-ground C data distribution. For each variable we report in parenthesis its relative importance.*

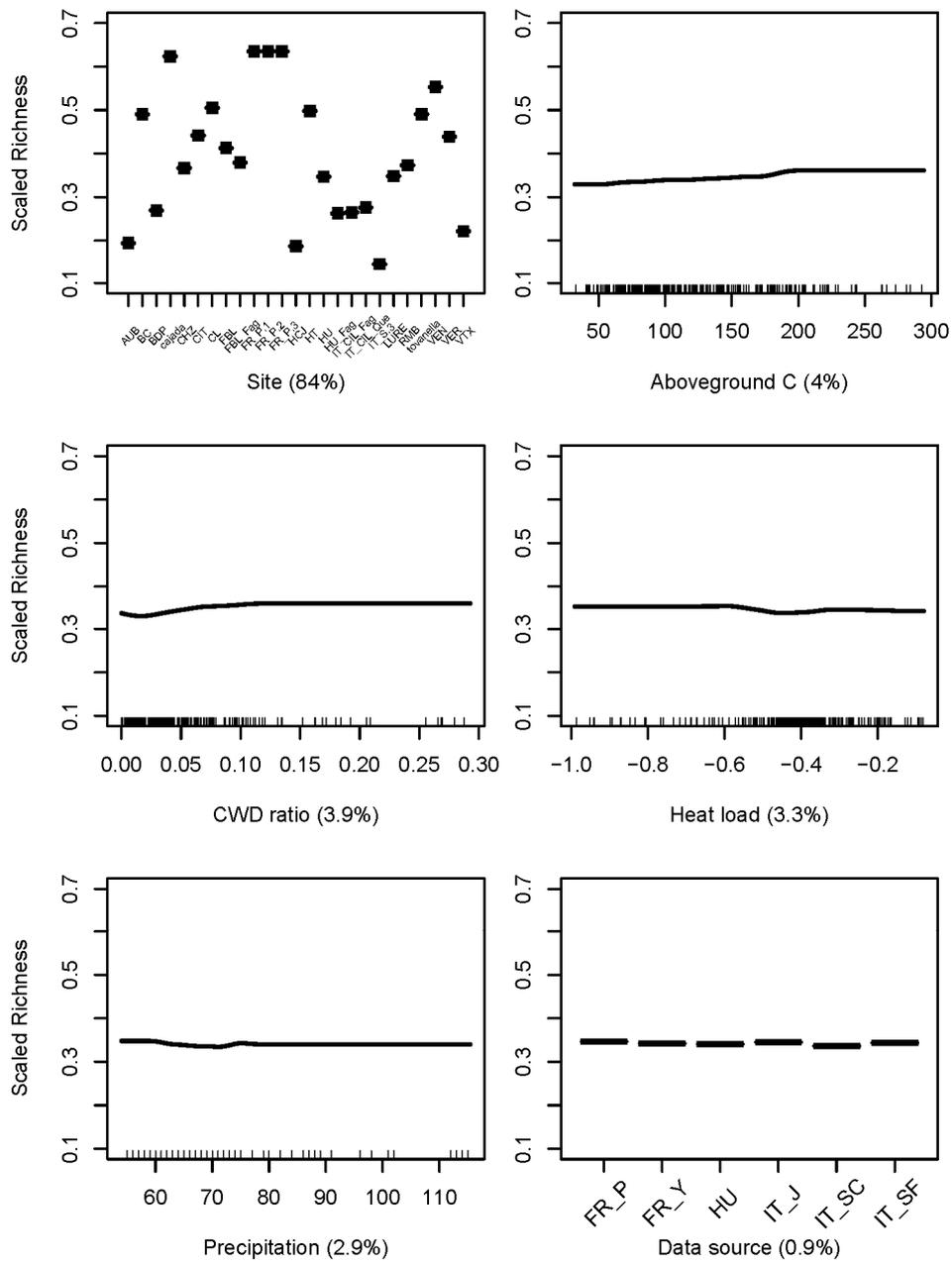

*Figure S2 – Partial dependency plots of the relationship between the scaled richness of birds and the six top-performing explanatory variables as modelled through a Boosted Regression Tree.*

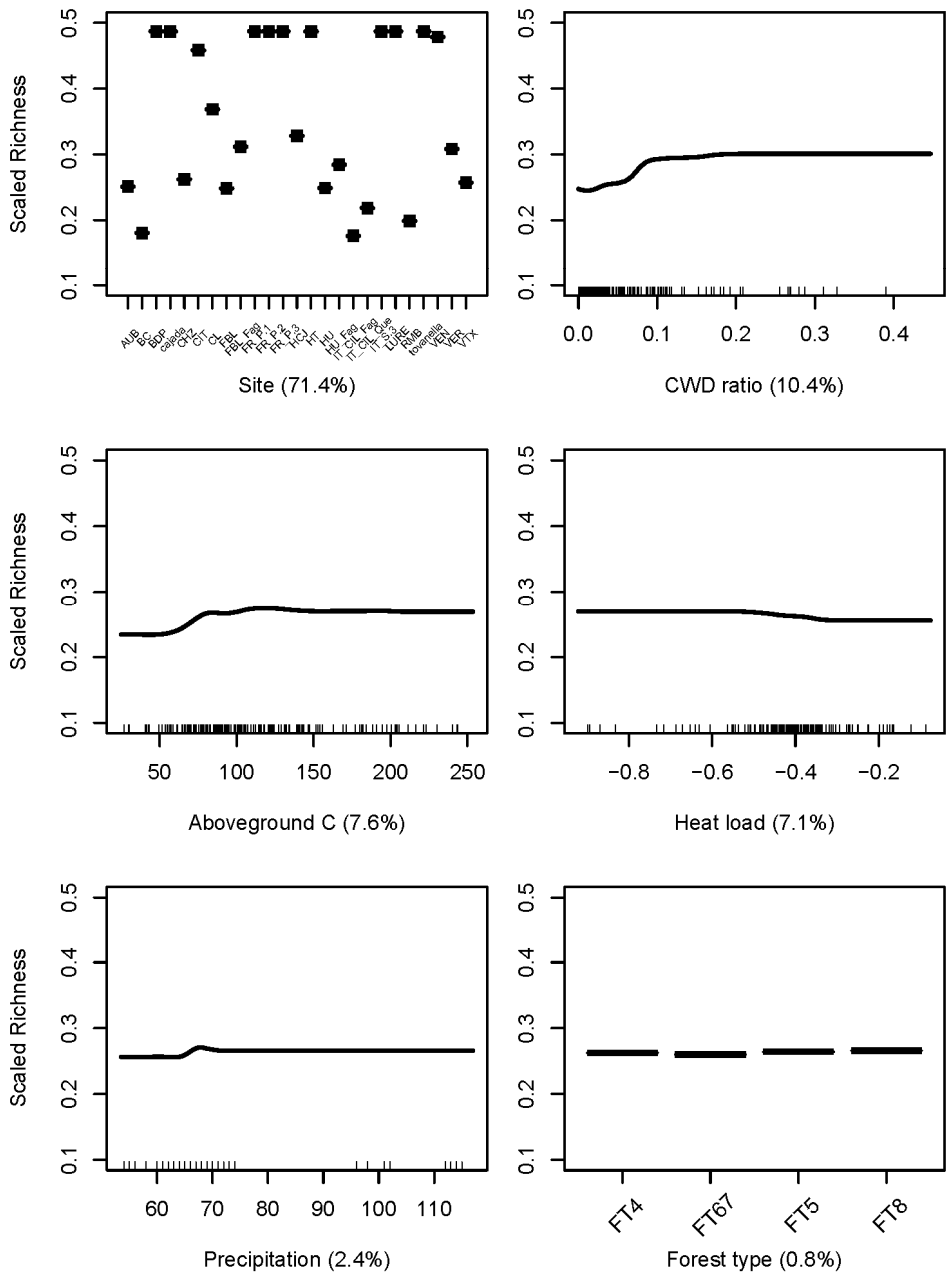

*Figure S3 – Partial dependency plots of the relationship between the scaled richness of bryophytes and the six top-performing explanatory variables as modelled through a Boosted Regression Tree.*

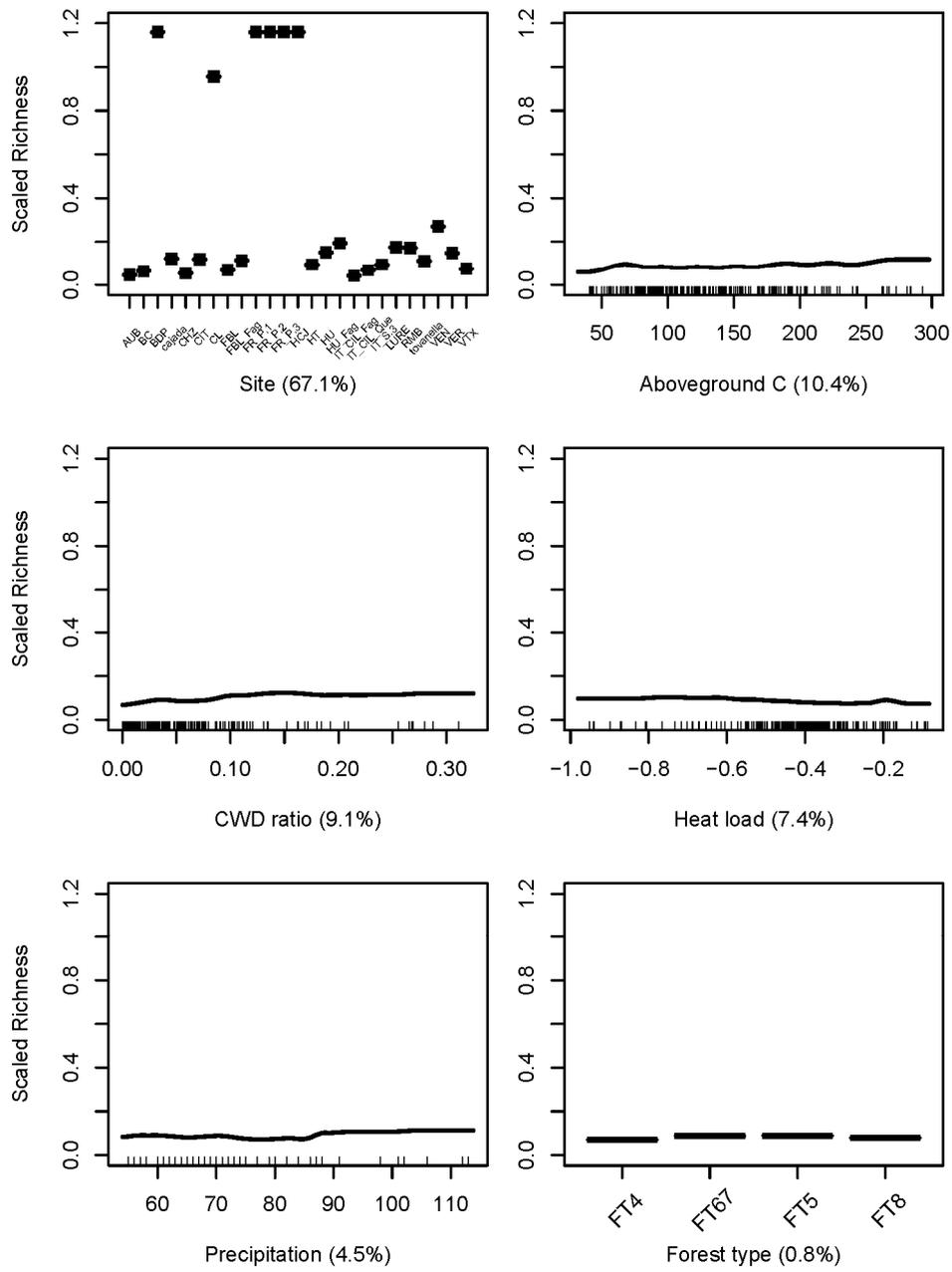

*Figure S4 - Partial dependency plots of the relationship between the scaled richness of fungi and the six top-performing explanatory variables as modelled through a Boosted Regression Tree.*

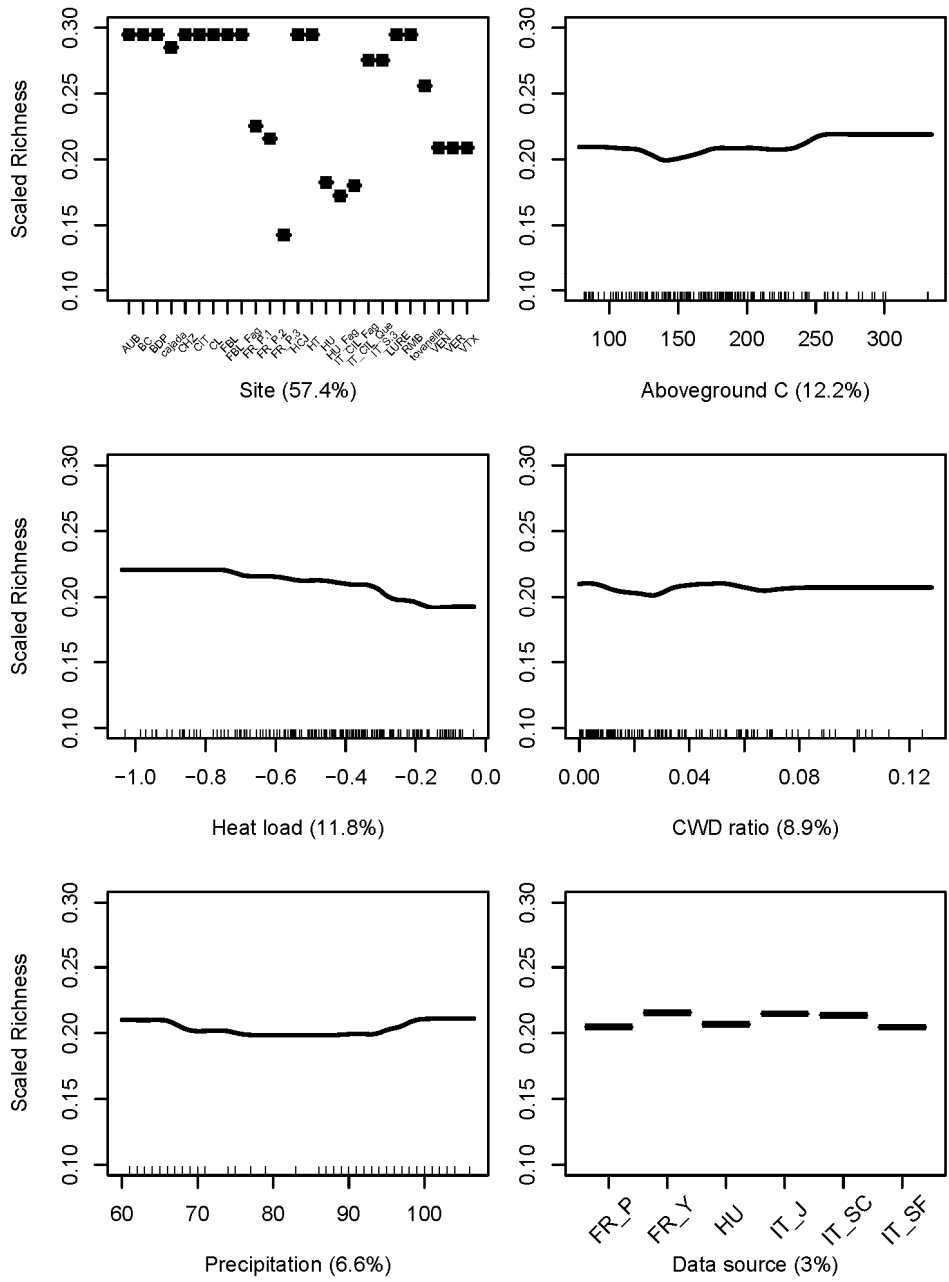

*Figure S5 - Partial dependency plots of the relationship between the scaled richness of lichens and the six top-performing explanatory variables as modelled through a Boosted Regression Tree.*

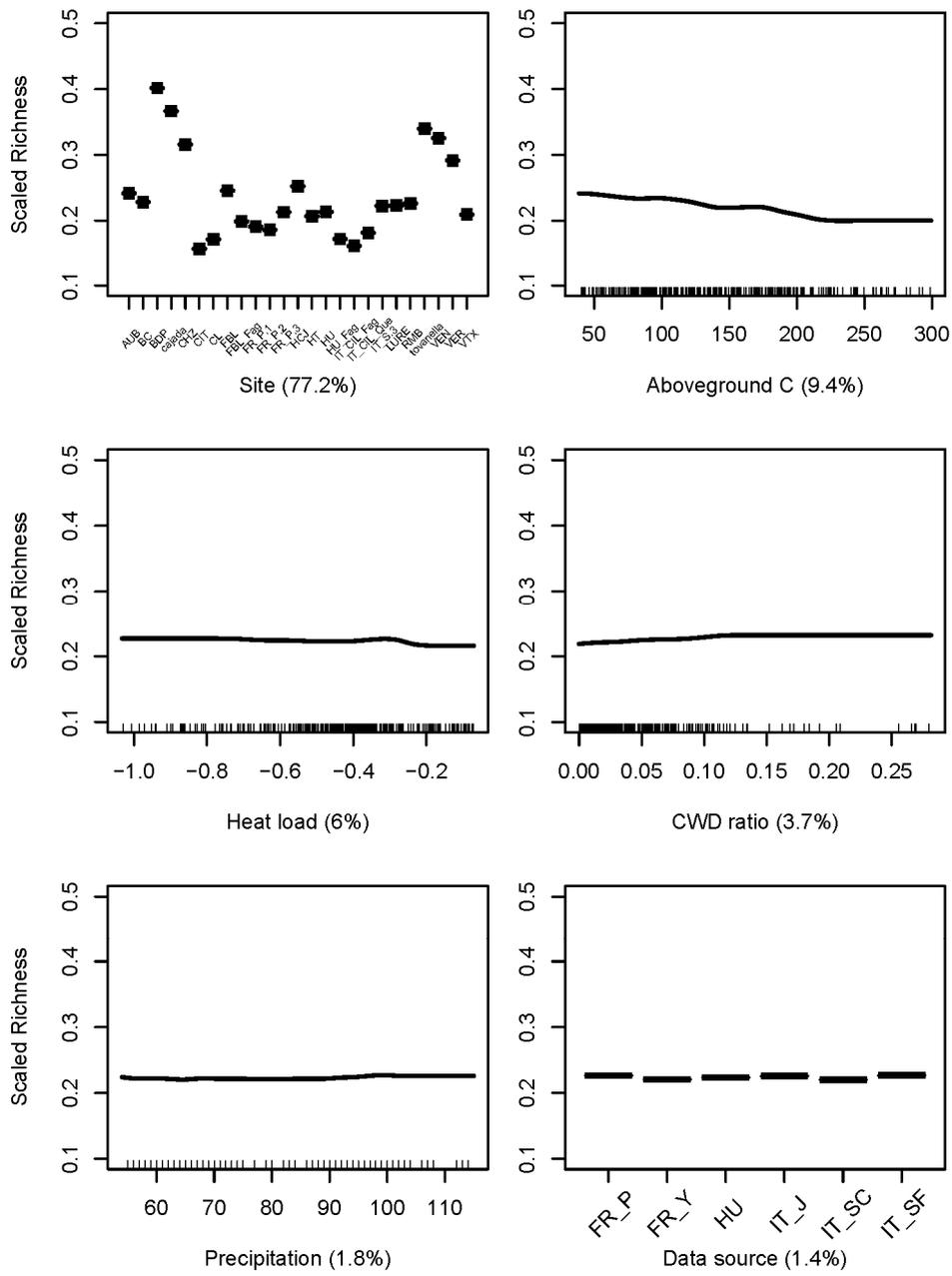

*Figure S6 - Partial dependency plots of the relationship between the scaled richness of plants and the six top-performing explanatory variables as modelled through a Boosted Regression Tree.*

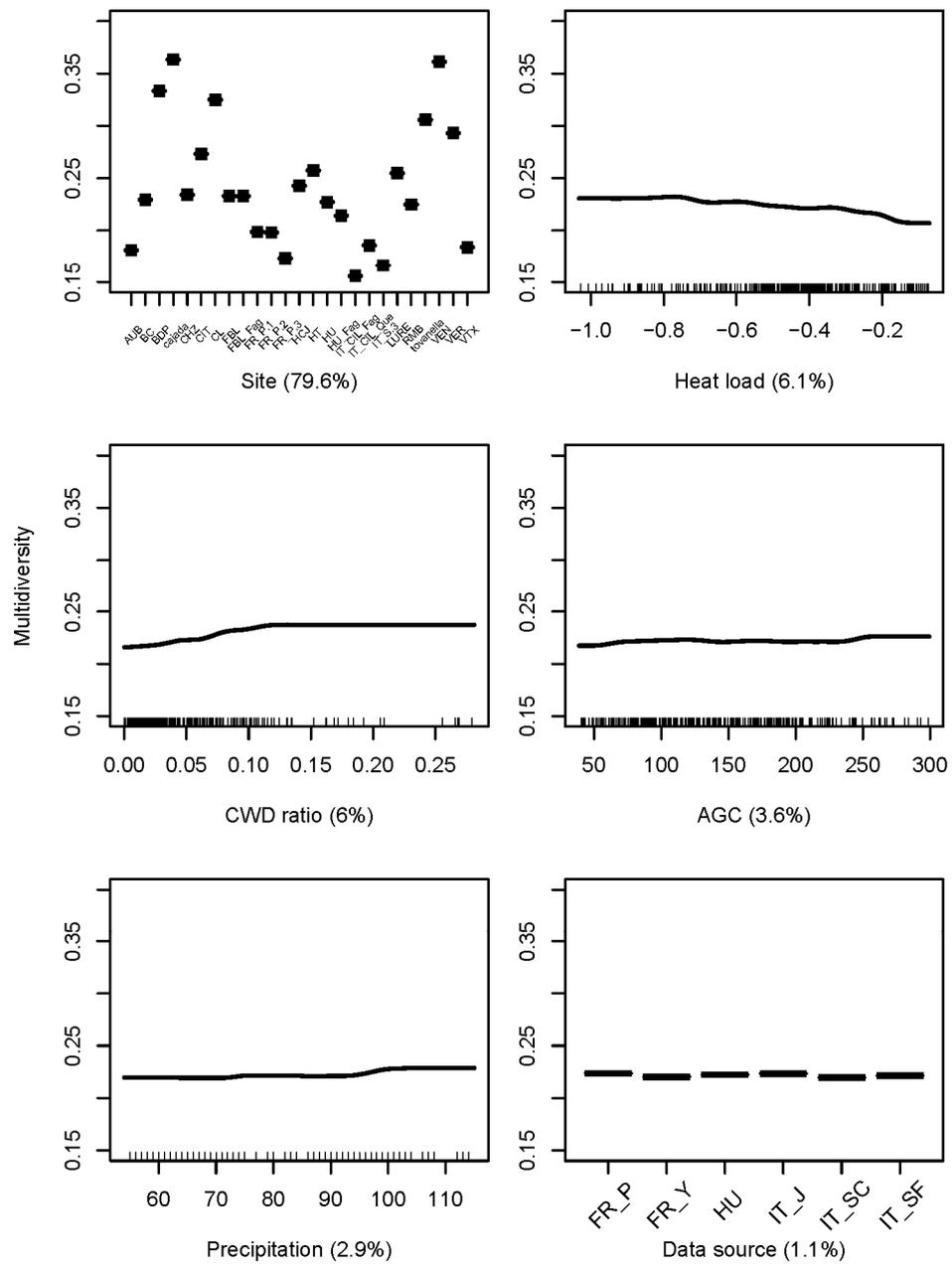

*Figure S7 – Partial dependency plots of the relationship between the scaled richness of multidiversity and the six top-performing explanatory variables as modelled through a Boosted Regression Tree.*

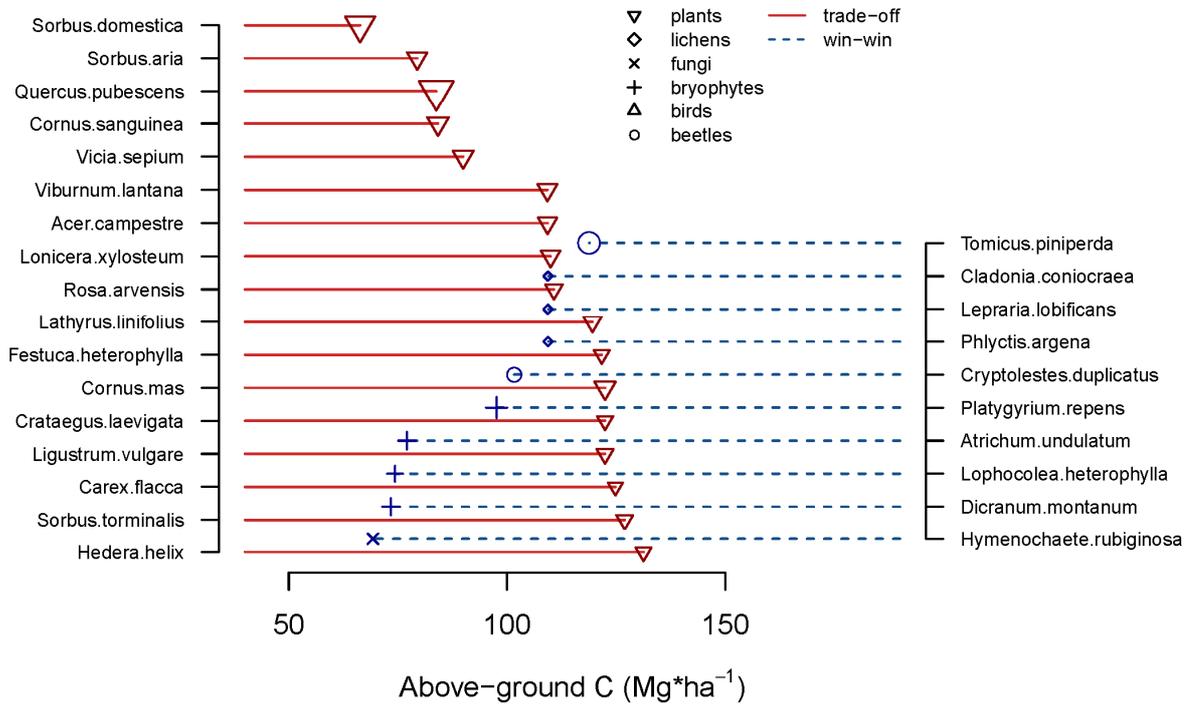

*Figure S8* – Diving board plot of species-specific change-points for pure and reliable indicator species in oak-dominated forests. For win-win species, the horizontal lines extend from the highest observed above-ground C value to the 5$^{th}$ percentile of the permuted distribution of change-points. For trade-off species the horizontal lines extend from the lowest observed above-ground C value to the 95$^{th}$ percentile. The size of the symbols is proportional to species' indicator value (z-scores). Only species having a level of purity and reliability higher than 0.95 were reported. We reported the limits of the confidence interval, as these conservatively indicate the point along the above-ground C gradient above (below) which the trade-off (win-win) species potentially decline (increase).

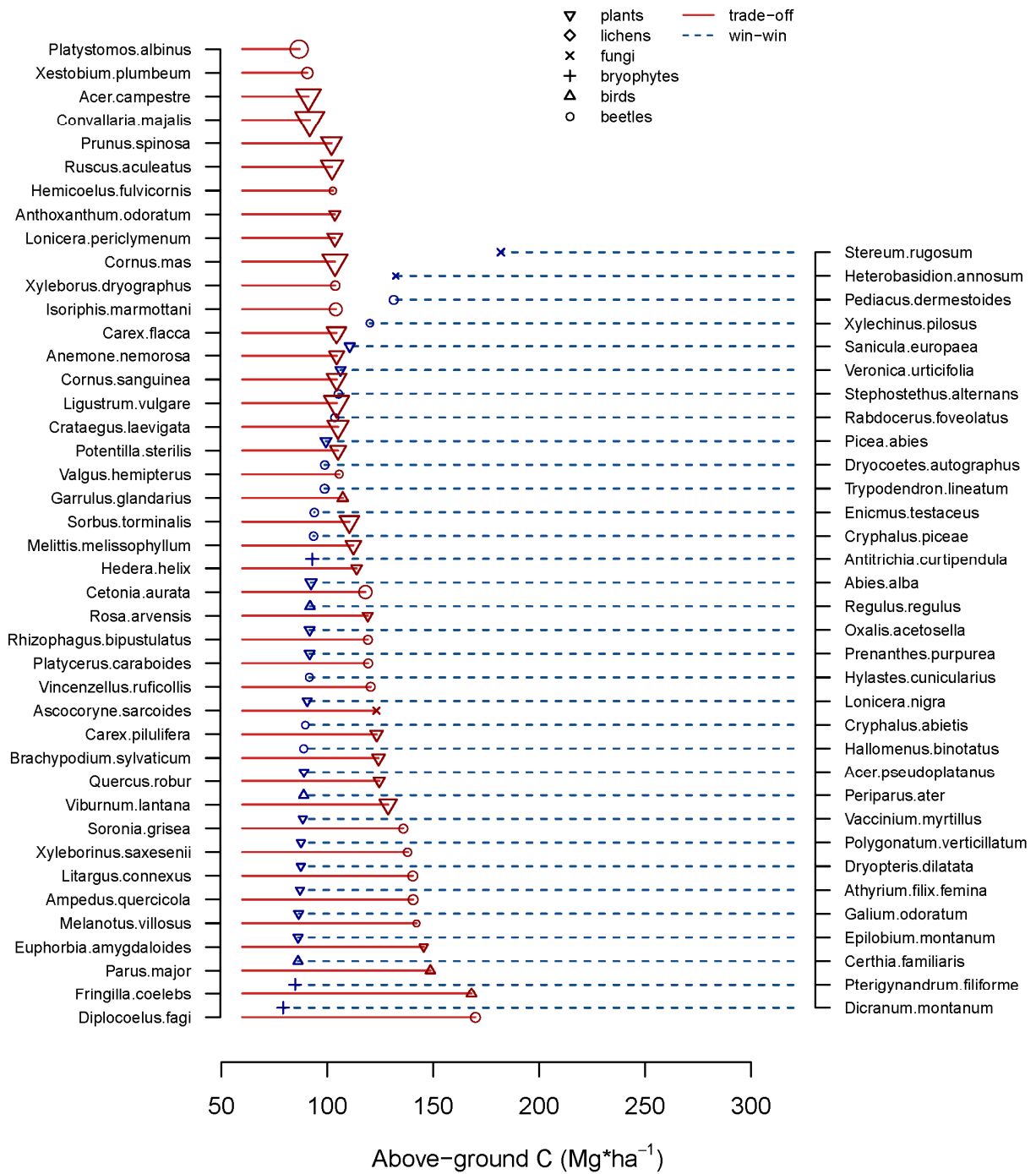

*Figure S9 – Diving board plot of species-specific change-points for pure and reliable indicator species in beech forests.*

*Symbols as in Figure S8*